\documentclass[hidelinks]{SBCbookchapter}
\usepackage[utf8]{inputenc}
\usepackage[T1]{fontenc}
\usepackage[brazil,english]{babel}
\usepackage{graphicx}

\usepackage{cite}
\usepackage{comment}
\usepackage{amssymb}

\author{Florian Willomitzer \\
Wyant College of Optical Sciences - University of Arizona}

\title{Synthetic Wavelength Imaging - \\ Utilizing Spectral Correlations for High-Precision Time-of-Flight Sensing}

\begin{document}

\maketitle

\section{Introduction} 
\label{FWillomitzer:sec:Introduction}

Optical three-dimensional (3D) imaging \index{3D imaging} and ranging techniques have been used in academia and industry for many years with great success. Fields of application include medical imaging, autonomous navigation, industrial inspection, forensics, or virtual reality. The great success of these techniques is no coincidence, as high-quality 3D object or scene representations contain a distinctive higher information content than simple 2D images: Compared to 2D images, 3D representations are invariant against translation and rotation of the object as well as variations in surface texture or external illumination conditions. This fact has made 3D imaging an established tool in optical metrology and, more recently, in computer vision. \\
Although the sheer number of available 3D sensing principles
is immense, existing approaches can be broadly categorized into
3 groups: (1) triangulation-based approaches \index{Triangulation} \index{Structured light} (including active stereo, passive stereo, or focus search) \cite{ Srinivasan84, Schaffer10, willomitzer20103d, Takeda83, schoenberger16, arold2014hand, WilloOE_17, ettl2013improved}, (2) reflectance-based approaches that measure the surface gradient \index{Photometric stereo} \index{Deflectometry} (including photometric stereo and deflectometry) \cite{Woodham80, Horn90, Willo20PMD, Knauer04, faber12, Wang21, Huang2018}, and (3) approaches that measure the "time of flight" (ToF) or travel distance of light. This last group includes interferometry\index{Time-of-Flight}\index{Interferometry} and so-called "ToF-cameras". We will see later that the 3rd group can be again subdivided into ToF imaging on \textit{smooth} and \textit{rough} surfaces, although this distinction will not be discussed in detail (see~\cite{GH22LAM} for more information). 

This book chapter focuses on \textit{interferometric multi-wavelength ToF imaging techniques} - a subset of the 3rd category. The chapter should serve as a gentle introduction and is intended for computational imaging scientists and students new to this fascinating topic. Technical details (such as detector or light source specifications) will be largely omitted. Instead, the similarities between different methods will be emphasized to "draw the bigger picture." \\

In computer vision, ToF cameras\index{Time-of-Flight}\index{Time-of-Flight!camera}\index{Time-of-Flight!imaging} have become increasingly popular over the last years. Current techniques either utilize pulsed waveform modulation / LIDAR~\cite{collis1970lidar, weitkamp2006lidar} (which has already been discussed in previous chapters of this book)  or continuous wave amplitude modulation (CWAM)~\cite{schwarte1997new, lange2001solid, foix2011lock}.  
CWAM ToF cameras illuminate the scene with a light source whose intensity
is modulated over time, e.g., as a sinusoid.
The detector pixels in these devices behave as homodyne receivers which accumulate a charge proportional to the phase
difference $\phi(\lambda_{mod})$ between the modulated emitted light, and the modulated irradiance received at the
sensor. For each pixel, the scene distance $z$ can then be estimated via

\begin{equation}
z = \frac{1}{2} \frac{\lambda_{mod} \cdot \phi(\lambda_{mod})}{2 \pi}
\label{FWillomitzer:eq:Phase2Depth}
\end{equation}

where $\lambda_{mod}$ is the modulation wavelength of the  temporal intensity modulation of the  light source. The depth precision $\delta z$ is  directly proportional
to the modulation wavelength $\lambda_{mod}$: the smaller the wavelength, the better the depth precision\index{Depth precision}\index{Time-of-Flight!depth precision}~\cite{lange2001solid, Li21}. 

\begin{equation}
\delta z \sim \lambda_{mod}
\label{FWillomitzer:eq:PrecMod}
\end{equation}

However, current limitations in silicon manufacturing technology~\cite{lange2001solid, schwarte1997new, li2018sh} restrict the minimal achievable modulation wavelengths for CWAM ToF cameras to roughly $\lambda_{mod} \gtrsim 3m$, which results in a depth precision not better than centimeters. This might be sufficient for simple object detection tasks (i.e., "is there an object or not?"), or rough depth estimations (i.e., "is the object in the foreground or background?") but  leaves much room for improvement for more sophisticated computer vision or metrology applications, where precise measurement of the 3D geometry is important (e.g., in medical imaging or industrial inspection). \\
Alternatives with higher precision reside on the other end of the spectrum of ToF principles: \textit{Optical interferometers}~\cite{Wyant15OSA, BornWolf}. As mentioned above, ToF cameras and optical interferometers are closely related. Both systems evaluate the difference in path length or travel time  between the signal reflected or scattered off the object under test and a reference signal. This difference manifests as a phase shift in the detected signal. Instead of using an intensity-modulated light source or a pulse, optical interferometry exploits the oscillations of the physical wave light field itself, meaning that it works with very small optical wavelengths down to several hundred $nm$. For well-adjusted interferometers, this leads to depth precisions better than $nm$ on specular surfaces, which by far surpasses the precision that can be reached with a ToF camera. However, this very high precision of single-wavelength interferometry systems comes with a tradeoff: The inability to measure rough surfaces. If coherent light is scattered off a rough surface, the optical phase of the backscattered light field is randomized. The resulting "speckle field" does seemingly not contain any information about the macroscopic optical path length anymore, and range information cannot be extracted from one single (monochromatic) speckle field.

In this book chapter, it will be discussed how techniques known from multi-wavelength interferometry\index{Multi-wavelength interferometry} can be applied to compensate for the problem of phase randomization in a speckle field which is formed when a rough surface is illuminated with coherent light. This becomes possible, as multi-wavelength interferometry exploits information from an additional modality: \textit{spectral diversity}, acquired via additional measurements at different optical wavelengths. It will be shown how to utilize this additional modality for "synthetic wavelength imaging" concepts, which eventually leads to the development of interferometric Time-of-Flight (ToF) cameras with $\mu m$-range precision for the measurement of macroscopic objects with rough surfaces~\cite{li2018sh, Li:19, Wu20, Li21}.  Subsequent,  the concept of "synthetic waves" will be combined with digital holography to develop a novel approach for Non-Line-of-Sight (NLoS) imaging, i.e., to image hidden objects around corners or through scattering media with sub-$mm$ resolution~\cite{ WilloCOSI19, Willo19Arx, Willo21Nat}.

\section{Synthetic Wavelength Imaging} 
\label{FWillomitzer:sec:SyntheticWavelengthImaging}

\begin{figure}[t]
\centering
\includegraphics[width=1\linewidth]{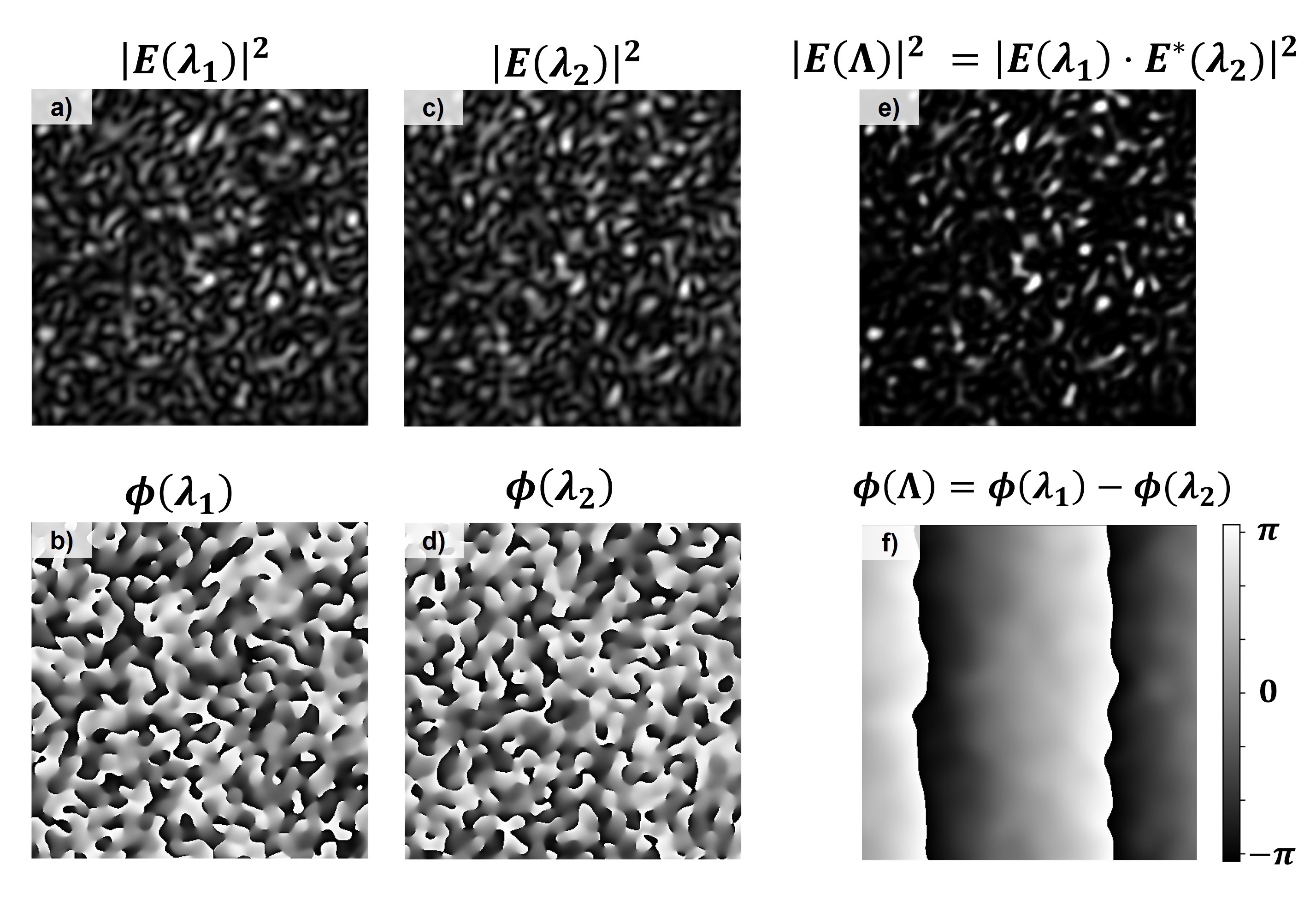}
\caption{Formation of the synthetic wave field, explained at the example of the measurement of a tilted planar rough surface (experiment): the surface is subsequently illuminated with a coherent beam at wavelength $\lambda_1 = 854.39nm$ and $\lambda_2 = 854.47nm$,  and the complex speckle fields $E(\lambda_1)$ and $E(\lambda_2)$ are captured with an interferometer. The intensity of both fields (a and c) shows the typical distribution of a speckle pattern,  and the phase of both fields (b and d) is randomized. The synthetic field $E(\Lambda)$ is formed (in this example) via mixing (Eq.~\ref{FWillomitzer:eq:Mixing}) and the synthetic wavelength calculates to $\Lambda \approx 9mm$. The microscopic path length variations cancel each other out, and the phase of the synthetic field (f) is not subject to phase randomization anymore. Macropsopic information about the surface (shape, tilt, etc.) can be extracted from (f).}
\label{FWillomitzer:fig:SW_Speckle}
\end{figure}

In the following, the term  "synthetic wavelength imaging" collectively\index{Synthetic wavelength}\index{Synthetic wavelength!imaging} groups imaging principles that use well-known techniques from optical metrology (such as interferometry or holography), but perform these techniques at a so-called "synthetic wavelength" instead of optical wavelengths. In many of the introduced methods, the synthetic wave field only "lives" in the computer, and the respective operations (such as propagation, superposition, etc.) are performed purely computationally. \\
As mentioned, synthetic wavelength imaging exploits concepts known from multi-wavelength interferometry, a technique that has been widely used in optical metrology (see exemplary references~\cite{ Cheng85, polhemus73, tiziani1996dual, PdG94,  falaggis2009multiwavelength, zhou2022review, PdG92}). For many implementations of multi-wavelength interferometry, the additional spectral information is utilized to increase the unambiguous measurement range for the interferometric measurement of optically smooth surfaces with large height variations. The resulting increase in dynamic range is available for the high-precision measurements of optical components or technical parts. It turns out, however, that  the utilization of spectral diversity\index{Spectral diversity} in interferometry has another significant benefit: It enables interferometric measurements of\index{Rough surface!optically} \textit{optically rough surfaces}. This is very surprising at the first glance: It was just discussed that illuminating an optically rough surface with coherent light at wavelength $\lambda_1$ produces a\index{Speckle}\index{Speckle!formation}\index{Speckle!field} speckle field, which is randomized in phase and intensity~\cite{GH04Enc, GoodmanSpeck} and optical path length information cannot be inferred. However, by probing the scene with a second wavelength $\lambda_2$, slightly different from the initial wavelength $\lambda_1$, additional information can be exploited. The basic process is explained in Fig.~\ref{FWillomitzer:fig:SW_Speckle}: The speckle field $E(\lambda_1)$, emerging from a tilted planar rough surface which is illuminated with a spatially and temporally coherent beam at wavelength $\lambda_1$ is  randomized in intensity and phase (see Fig.~\ref{FWillomitzer:fig:SW_Speckle}a,b)\index{Synthetic wavelength!formation}. Illuminating the surface with a second beam at another wavelength $\lambda_2$ produces a second speckle field $E(\lambda_2)$ (see Fig.~\ref{FWillomitzer:fig:SW_Speckle}c,d). If the two beams at $\lambda_1$ and $\lambda_2$ originate from the same source (e.g., from the same fiber tip of the illumination unit), the respective fields \textit{undergo the exact same path length variations} before reaching the detector. For closely spaced wavelengths $\lambda_1$ and $\lambda_2$ this results in two speckle fields with a largely similar intensity distribution (see Fig.~\ref{FWillomitzer:fig:SW_Speckle}a,c) - the speckle fields are \textit{'spectrally correlated'}~\cite{GoodmanSpeck}. \\
The two complex valued speckle fields $E(\lambda_1)$ and $E(\lambda_2)$ incident on the detector can be expressed as

\begin{equation}
    E(\lambda_1) = A_1 \cdot e^{i \phi(\lambda_1)}~~, ~~
    E(\lambda_2) = A_2 \cdot e^{i \phi(\lambda_2)}~~,
\end{equation}

with $A_1, A_2$ being the respective amplitudes and $\phi(\lambda_1), \phi(\lambda_2)$  the respective (speckled) phase maps at the given wavelengths $\lambda_1, \lambda_2$. After measuring and storing $E(\lambda_1)$ and $E(\lambda_2)$, a so-called  "synthetic field" $E(\Lambda)$ can be calculated. One possibility to calculate $E(\Lambda)$ is \textit{computational mixing} of $E(\lambda_1)$ and $E(\lambda_2)$

\begin{equation}
    {E(\Lambda) = E(\lambda_1)\cdot E^*(\lambda_2)} \\
		{= A_1 A_2 \cdot e^{i(\phi(\lambda_1) - \phi(\lambda_2))} = A_1 A_2 \cdot e^{i\phi(\Lambda)}} ~~,
		\label{FWillomitzer:eq:Mixing}
     \end{equation}
     
where $E^*(\lambda_2)$ denotes the complex conjugate of $E(\lambda_2)$. Other options to calculate the synthetic field can be found in~\cite{li2018sh, Li21}. It can be seen in Eq.~\ref{FWillomitzer:eq:Mixing} that the phase of $E(\Lambda)$ is composed of the \textit{difference of the two optical phase maps}. This means that the synthetic field  $E(\Lambda)$ contains information that is on the order of a "synthetic wavelength" $\Lambda$, which is the beat wavelength of $\lambda_1$ and $\lambda_2$: \index{Synthetic wavelength}

\begin{equation}
\Lambda = \frac{\lambda_1 \cdot \lambda_2}{|\lambda_1 - \lambda_2|}  
\label{FWillomitzer:eq:SWL}
\end{equation}

With a \textit{closely spaced} selection of $\lambda_1$ and $\lambda_2$, the synthetic wavelength can be chosen orders of magnitudes larger than the two optical wavelengths or the microscopic roughness \footnote{For the sake of brevity and a gentle introduction, an exact definition of the term "surface roughness" is omitted at this early point of the chapter. The topic will be described in detail in section~\ref{FWillomitzer:sec:Limits}, when the theoretical limits of the introduced methods are discussed.} of the surface. For example, the selection of the two optical wavelengths $\lambda_1 = 550nm$ and $\lambda_2 = 550.1nm$ leads to a synthetic wavelength of $\Lambda \approx 3mm$. For surfaces with roughness $ \ll \Lambda $ the fields $E(\lambda_1)$ and $E(\lambda_2)$ are \textit{highly correlated}~\cite{GoodmanSpeck}. Remaining phase fluctuations in the synthetic phase map $\phi(\Lambda)= \phi(\lambda_1) - \phi(\lambda_2)$ are negligible and $\phi(\Lambda)$ becomes largely equivalent to the phase map \textit{that would be measured for a source that emits electromagnetic radiation at the much larger wavelength $\Lambda$}, meaning that it displays no speckle artifacts (see Fig.~\ref{FWillomitzer:fig:SW_Speckle}f). This is possible, since the phase perturbations, imparted for each field by the microscopic path length variations, cancel each other out and information about the macroscopic path length variation is now visible in the resulting phase map $\phi(\Lambda) $. It should be emphasized that this is only the case for the \textit{phase map} and not the intensity term $|E(\Lambda)|^2 = |E(\lambda_1) \cdot E^*(\lambda_2)|^2$, which still displays the spatial structure of the two speckle fields at the optical wavelengths $\lambda_1$ and $\lambda_2$ (see Fig.~\ref{FWillomitzer:fig:SW_Speckle}e).

\begin{figure}[t]
    \centering
    \includegraphics[width=0.8\linewidth]{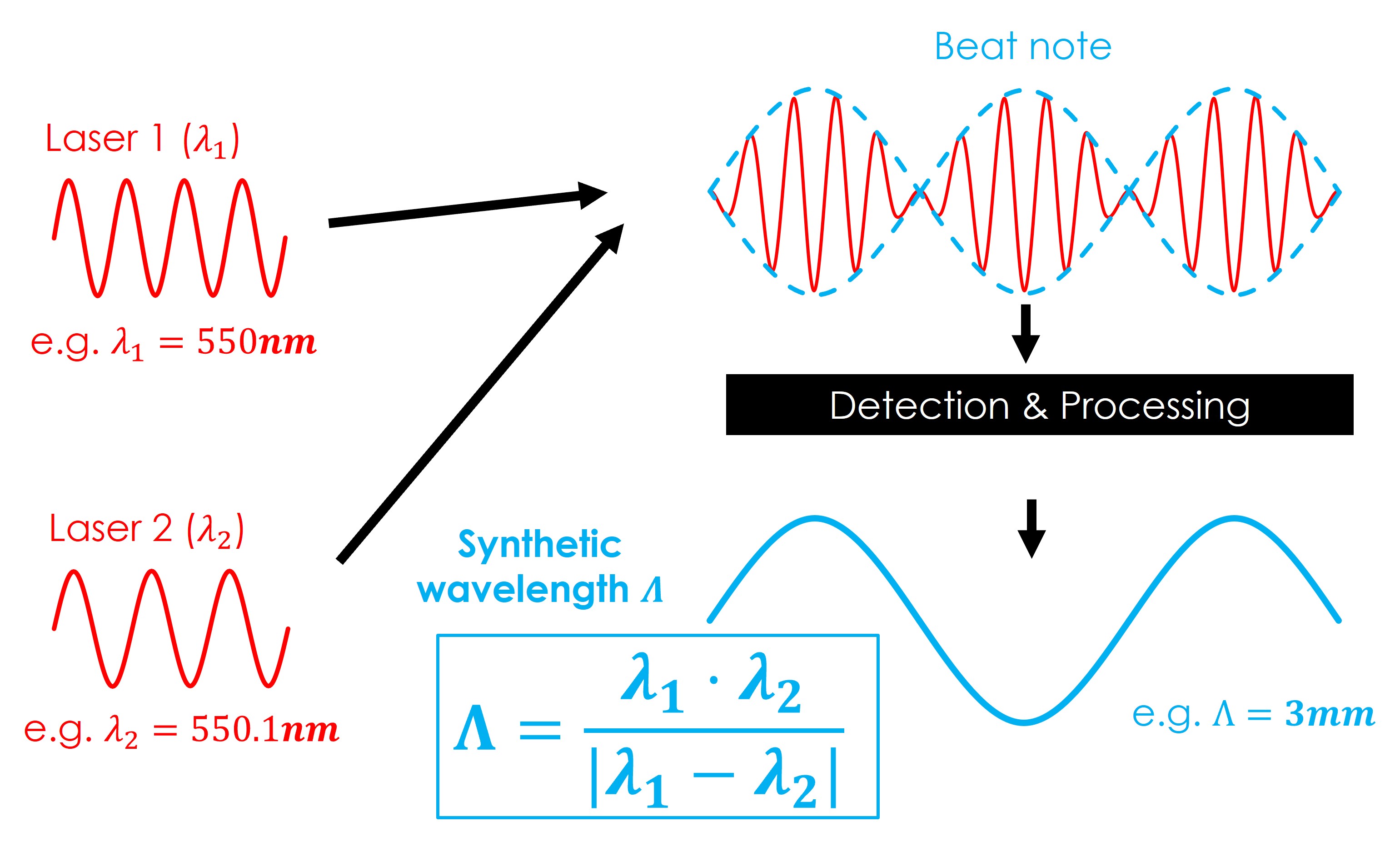}
    \caption{Synthetic wavelength imaging exploits spectral diversity, i.e., encoding at different optical illumination wavelengths. The superposition of two or more optical fields at slightly different wavelengths (in this example, $\lambda_1 = 550nm $ and $\lambda_1 = 550.1nm $) produces a high-frequency wave with a low-frequency beat note. The beat wave can be  isolated by special detection and processing methods and forms the "synthetic wave," where the synthetic wavelength (in this example $\Lambda = 3mm$) is calculated by Eq. ~\ref{FWillomitzer:eq:SWL}. For the majority of results shown in this book chapter, the synthetic field is obtained via mixing of the two optical fields (see Eq.~\ref{FWillomitzer:eq:Mixing}). }
    \label{FWillomitzer:fig:SW_Basic}
\end{figure}

Obtaining the largely speckle-free synthetic phase map $\phi(\Lambda)$ is the first step of further computational processing. To evaluate the final result,  synthetic wavelength imaging procedures often utilize well-known methods of computational optics (such as computational wave propagation)  and perform these methods "at the synthetic wavelength." This is possible, as  the computationally constructed synthetic wave behaves much like an optical wave~\cite{Willo21Nat, Willo21NatSupp}, and hence can be processed using the same algorithms. \\
In the following, two examples of this striking similarity are introduced: \textit{Synthetic Wavelength Interferometry} for the precise "ToF" 3D measurement of macroscopic objects with rough surfaces~\cite{li2018sh, Li:19, Wu20, Li21}, and \textit{Synthetic Wavelength Holography} which can be used to image hidden objects around corners or through scattering media~\cite{WilloCOSI19, Willo19Arx, Willo21Nat}.   In both examples, the calculation of the synthetic field $E(\Lambda)$ helps to circumvent the deleterious effects of speckle arising at the optical wavelengths. However, it should be emphasized that the fields at the optical wavelengths are still the carrier of the information, and inexpensive sensors for visible light (no mm-wave detectors) can be used for their detection.  For the sake of brevity, the methods are mainly introduced with work performed by the author's research group.

\section{Synthetic Wavelength Interferometry} 

Interferometric\index{Synthetic wavelength}\index{Synthetic wavelength!interferometry} imaging methods obtain 3D shape information by measuring the phase of a light field reflected off the surface under test with respect to a reference field~\cite{Wyant15OSA, HechtOptics}.
As common for principles measuring the standoff distance of an object via optical path length differences (such as interferometry or ToF imaging), the unique measurement range\index{Unique measurement range} $\Delta Z$ is defined by 1/2 of the wavelength $\lambda_{mod}$ at which the signal is modulated ($\Delta Z = \frac{\lambda_{mod}}{2}$). The distance value $z$ is calculated by Eq.~\ref{FWillomitzer:eq:Phase2Depth} after measuring the respective phase map $\phi(\lambda_{mod})$. As discussed, CWAM ToF cameras exploit intensity modulation of an (in most cases) incoherent light source~\cite{schwarte1997new, lange2001solid}, while optical interferometry utilizes the modulation of the physical wave light field itself, meaning that $\lambda_{mod} = \lambda$ is within or close to the visible wave band. This reveals an interesting trade-off: The extremely high precision of optical (single-wavelength) interferometry  comes at the price of an extremely small unique measurement range $\Delta Z = \frac{\lambda}{2}$ (several hundred $nm$ for visible light), and phase wrapping occurs for surface height variations exceeding $\Delta Z$. Phase unwrapping methods can be used to extend the unique measurement range $\Delta Z$ and to disambiguate the captured phase map $\phi(\lambda)$. The probably most prominent example is\index{Phase unwrapping}\index{Phase unwrapping!multi-frequency}  multi-frequency phase unwrapping~\cite{huntley1993temporal}, which requires the measurement of a second phase map at a different wavelength. \\
Although the exact details will be discussed later, it should be briefly mentioned here that synthetic wavelength imaging can also be seen as a special form of multi-frequency phase unwrapping. This is because the technique is able to disambiguate the random phase fluctuations in speckled interferograms. 
The microscopic path length variations in the two optical interferograms at $\lambda_1$ and $\lambda_2$ cancel each other out, and the final synthetic phase map $\phi(\Lambda)$ contains only information about the macroscopic path length variations on the order of the synthetic wavelength $\Lambda$. As discussed, $\phi(\Lambda)$ is virtually speckle-free if the distance between the two optical wavelengths $\lambda_1$ and $\lambda_2$ is chosen sufficiently small (i.e., the synthetic wavelength $\Lambda$ is sufficiently large). In this case, the surface height can be simply determined by

\begin{equation}
z = \frac{1}{2}\frac{\Lambda \cdot \phi(\Lambda)}{2 \pi}~~.
\end{equation}

Note that this expression is equivalent to Eq.~\ref{FWillomitzer:eq:Phase2Depth}, with the synthetic wavelength being now the utilized modulation wavelength. It should also be emphasized that surface measurements at the synthetic wavelength do not exclude the possibility of phase wrapping. Depending on the ratio between surface height variations and unique measurement range $\Delta Z = \frac{\Lambda}{2}$, the synthetic phase map $\phi(\Lambda)$ could be wrapped. In analogy to classical interferometry, unwrapping can be performed. Besides the described multi-frequency phase unwrapping, which now would utilize a second \textit{synthetic} phase map acquired at a different synthetic wavelength~\cite{Li21}, other unwrapping methods could be applied as well. Examples are spatial phase unwrapping~\cite{WilloDiss_19}, or data-driven phase unwrapping procedures~\cite{wang2019one, yin2019temporal}. 

Synthetic wavelength interferometry is a long-known and well-established principle in optical surface metrology. Over the years, many different flavors of the synthetic wavelength idea have been published. The exemplary references~\cite{PdG92, Dandliker:88, Fercher:85, Vry:86, Li21, kotwal2022swept, zhou2022review} should give the reader an idea about the diversity of approaches and their applications. While the method of processing the synthetic interferograms  is fairly similar to most approaches, variations exist  in how the complex-valued speckle fields at the two or more optical wavelengths  are acquired. Common procedures include simultaneous, and sequential capturing of $E(\lambda_1)$ and $E(\lambda_2)$, and rely on, e.g., spatial heterodyning, frequency heterodyning, or phase shifting of the reference beam with respect to  the object beam. The latter can be performed, e.g., by moving a mirror in the reference arm. An overview of different methods is given in~\cite{zhou2022review, kreis1997methods, kim2010principles}.

\begin{figure}[b!]
\centering
\includegraphics[width=1\linewidth]{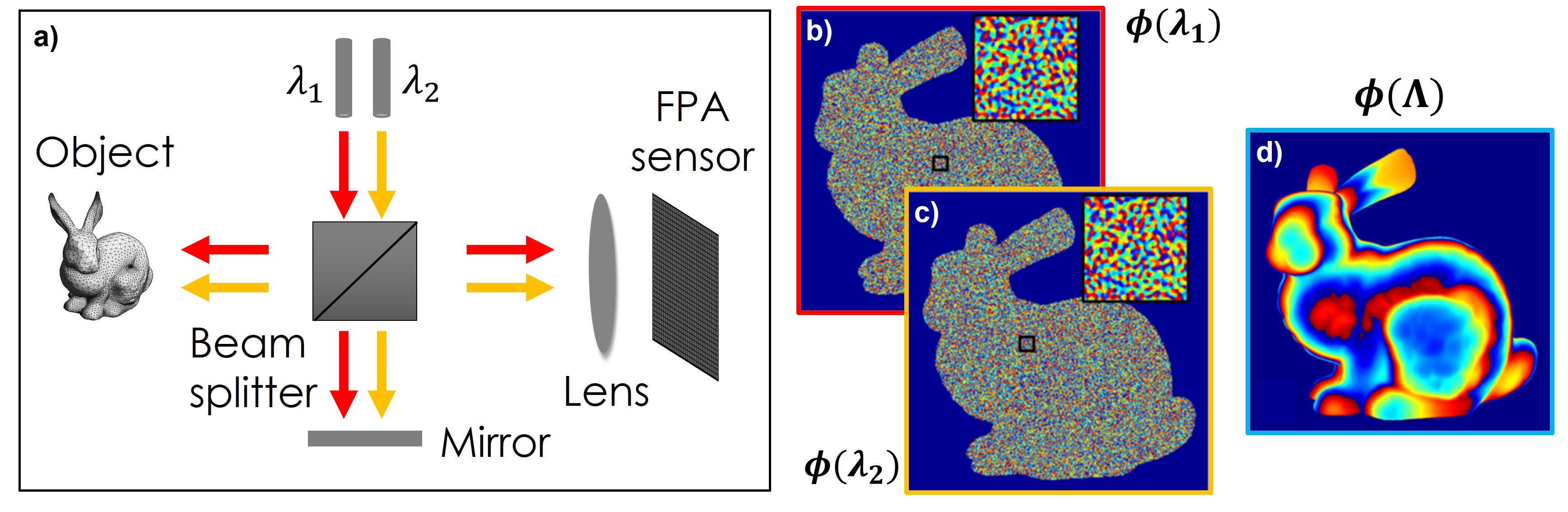}
\caption{High-precision ToF imaging with synthetic wavelength interferometry (simulation): a) Schematic setup of a dual-wavelength Michelson interferometer, which serves as "high-precision ToF camera". b) and c) Phase maps $\phi(\lambda_1)$ and $ \phi(\lambda_2)$ acquired at the two optical wavelengths. Both phase maps are subject to heavy phase randomization. d) Resulting speckle-free synthetic phase map $\phi(\Lambda)$, which allows for the extraction of depth information. }
\label{FWillomitzer:fig:SWI_Bunny}
\end{figure}

The remainder of this section focuses on a very specific application of synthetic wavelength interferometry which recently has been investigated by the author's research group: The precise 3D acquisition of macroscopic objects with rough surfaces for the specific application to problems in computer vision, medical imaging, automotive, or virtual reality. Or, in other words: \textit{Utilizing synthetic wavelength interferometry to build a "high-precision ToF camera."} As before, explanations will be kept to a high level, and technical details will be largely omitted. Further details can be found in~\cite{Li21, li2018sh, Li:19, Wu20}.

Figure~\ref{FWillomitzer:fig:SWI_Bunny}a  displays the schematic setup of the\index{High precision ToF camera}\index{Time-of-Flight} "high precision ToF camera," which consists of a  dual-wavelength Michelson interferometer\index{Interferometer!Michelson}, equipped with a lens that  images the object onto a high-resolution pixel array of a camera (herein  referred to as "focal plane array" (FPA) sensor). The wavelength of at least one laser in the setup is tunable so that different wavelength spacings $|\lambda_1 - \lambda_2|$, and hence different synthetic wavelengths $\Lambda$ can be realized. Amongst the various detection methods mentioned above, the author's research group has predominantly used heterodyne\index{Interferometer!heterodyne}~\cite{Fercher:85} and superheterodyne\index{Interferometer!superheterodyne}~\cite{Dandliker:88} detection, by utilizing the frequency modulation of additional acousto-optic modulators (AOMs) integrated with a fiber-based setup. To the best of its knowledge, the group was the first team who paired heterodyne/superheterodyne interferometry with high-resolution FPA detectors for full-field 3D measurements~\cite{Li21}. Raster-scanning-based superheterodyne interferometers equipped with a single-pixel detector have been studied as well~\cite{li2018sh}. \\
The simulated results in Fig.~\ref{FWillomitzer:fig:SWI_Bunny}b,c display exemplary phase maps acquired from  a macroscopic object with optically rough surface (bunny). As discussed before, it can be seen that $\phi(\lambda_1)$ and $ \phi(\lambda_2)$ are subject to heavy phase randomization due to speckle, while $\phi(\Lambda) $  (Fig.~\ref{FWillomitzer:fig:SWI_Bunny}d) is virtually speckle-free. In the realized example of an FPA-based heterodyne interferometer, the fields at both optical wavelength $E(\lambda_1), E(\lambda_2)$ are acquired   in a sequential fashion to form the synthetic field $E(\Lambda)$ via Eq.~\ref{FWillomitzer:eq:Mixing}.

\begin{figure}[h!]
\centering
\includegraphics[width=1\linewidth]{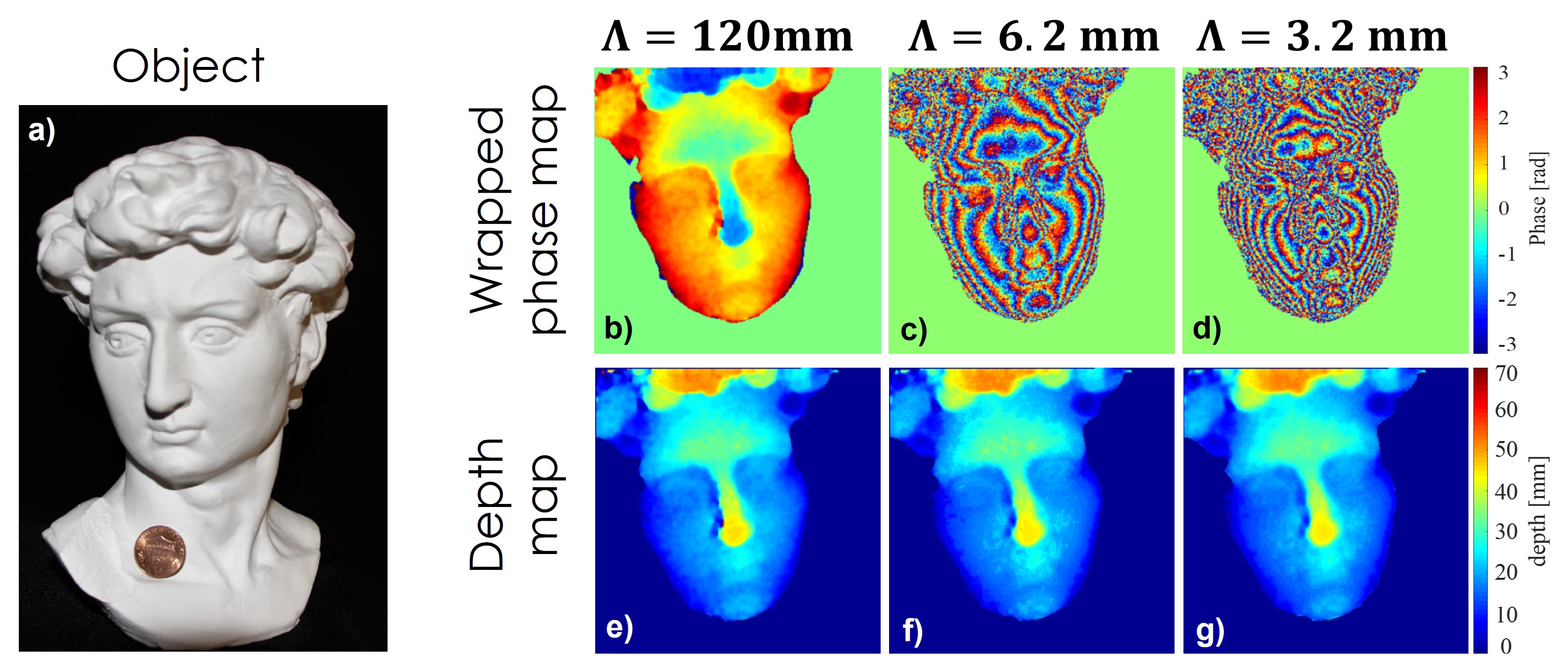}
\caption{Measurement of a plaster bust with a focal plane array (FPA) -based superheterodyne interferometer (experiment)~\cite{Li21}. a) Image of the bust with US penny for size comparison. b-d) Measured synthetic phase maps for different synthetic wavelengths ($120mm$, $6.2mm$, and $3.2mm$). e-g) Respective depth maps after phase unwrapping.}
\label{FWillomitzer:fig:SWI_Bust1}
\end{figure}

The demonstrated realization of the FPA-based superheterodyne interferometer~\cite{Li21} utilizes simultaneous illumination with both lasers  and is able to measure the synthetic phase map $\phi(\Lambda)$ directly - however, at the cost of $\geq 3$ phase shifts of the AOM modulated signal\index{Synthetic wavelength!measurements}.
Figure~\ref{FWillomitzer:fig:SWI_Bust1}b-d shows full-field synthetic phase measurements of a plaster bust, captured with the tunable dual-wavelength superheterodyne interferometer~\cite{Li21}. Synthetic phase maps acquired for different synthetic wavelengths are shown, and the field of view is approximately $10cm \times 10cm$. It can be seen that the phase maps at smaller synthetic wavelengths are subject to serious phase wrapping. After unwrapping (in the particular case  of Fig.~\ref{FWillomitzer:fig:SWI_Bust1} via multi-frequency phase unwrapping), the respective depth maps can be calculated (see Fig.~\ref{FWillomitzer:fig:SWI_Bust1}e-g). As expected, the noise level decreases for smaller synthetic wavelengths.

\begin{figure}[t]
\centering
\includegraphics[width=1\linewidth]{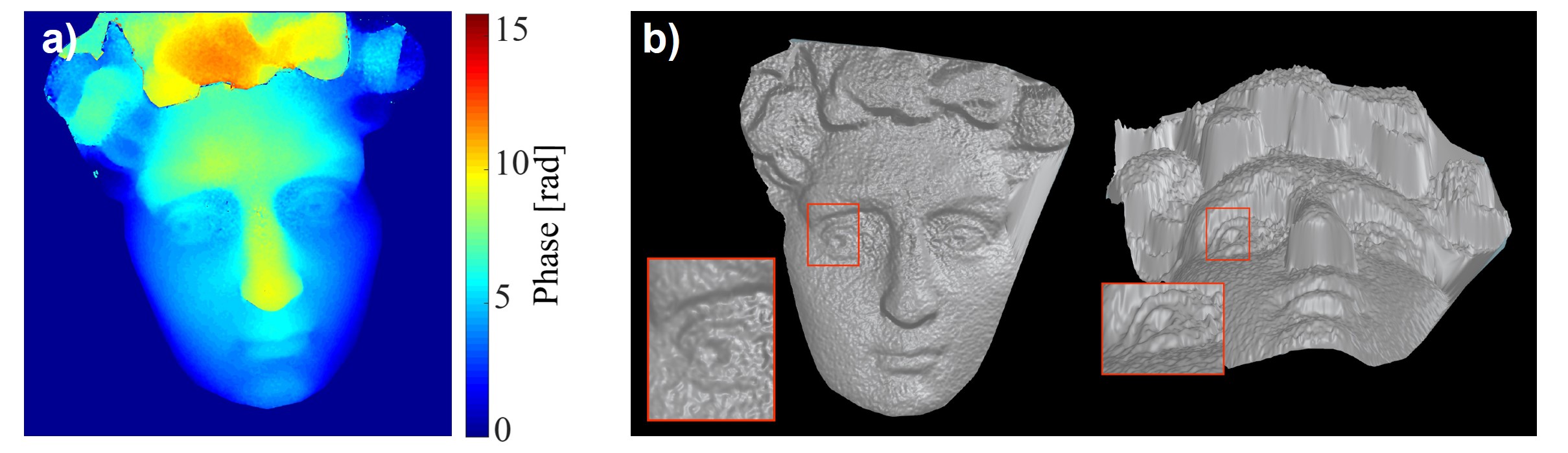}
\caption{Measurement of a plaster bust with a flutter-shutter camera-based tunable dual-wavelength heterodyne interferometer (experiment)~\cite{Li21}. The measurement was captured at a synthetic wavelength of $\Lambda \approx 43mm$ and the depth precision was evaluated to $\delta z < 380 \mu m$. a)~Measured phase map. b) Calculated 3D model shown from two different perspectives.}
\label{FWillomitzer:fig:SWI_Bust2}
\end{figure}

Figure~\ref{FWillomitzer:fig:SWI_Bust2} shows the 3D model of the same plaster bust, now acquired with a flutter-shutter camera-based tunable dual-wavelength heterodyne interferometer~\cite{Li21}. The 3D model was captured at a synthetic wavelength of $\Lambda \approx 43mm$, and the precision of the measurement was evaluated to $\delta z < 380 \mu m$. Measurements at smaller synthetic wavelengths reach higher precision. A comprehensive precision evaluation of the method is given in~\cite{Li21}, and the related fundamental limits are discussed in section~\ref{FWillomitzer:sec:Limits} of this chapter. 

\section{Synthetic Wavelength Holography} 

\begin{figure}[b!]
\centering
\includegraphics[width=1\linewidth]{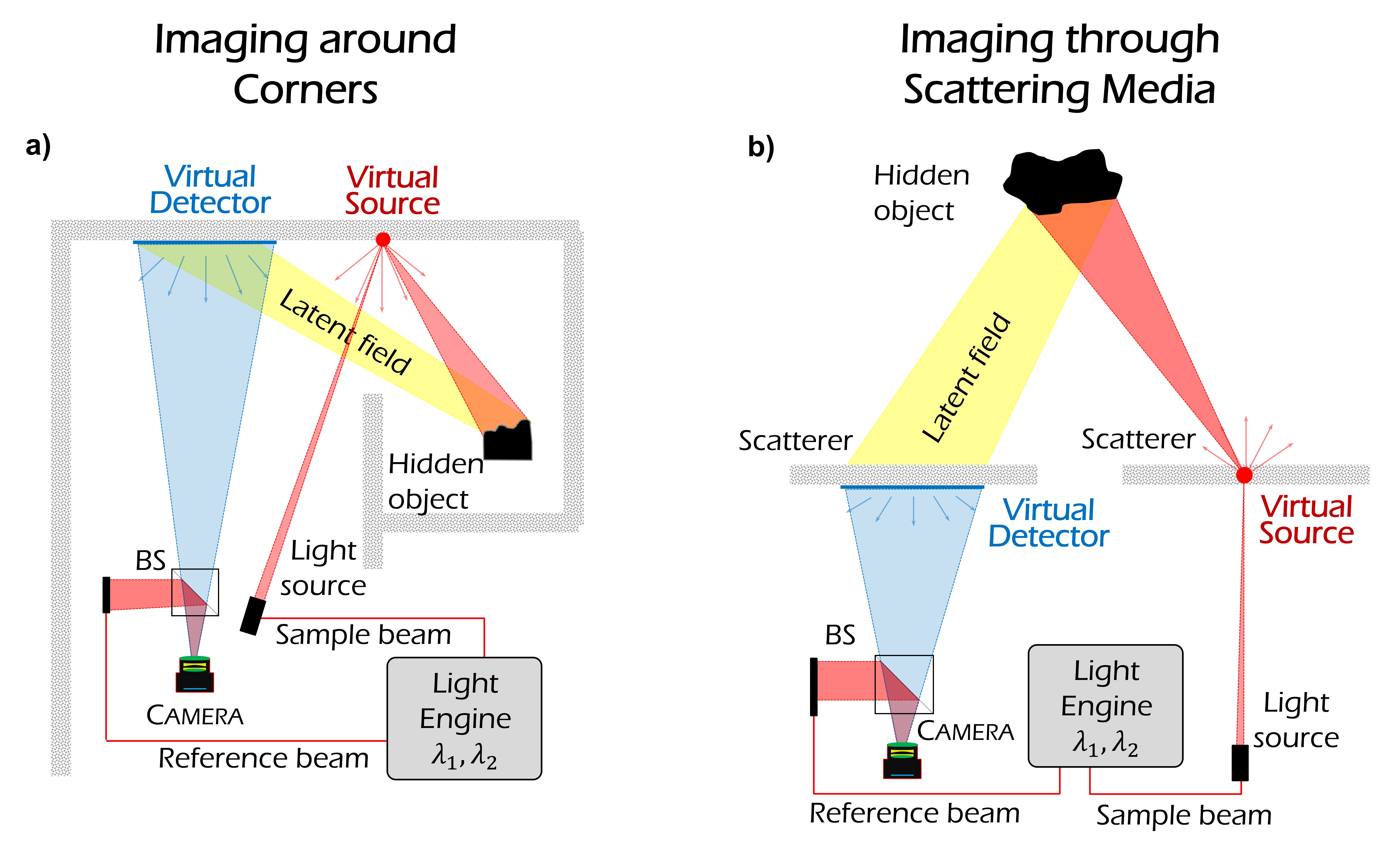}
\caption{Schematic setups for NLoS imaging around corners (a) and NLoS imaging through scatterers (b) with synthetic wavelength holography: A spot on the wall/scatterer (the "virtual source" VS) is illuminated by the sample beam. The VS scatters light towards the hidden object.
A small fraction of the light incident on the object is scattered back to the wall/scatterer, where it hits the "virtual detector" (VD). By imaging the VD with a camera, a  synthetic hologram is captured at the VD surface. Details about the light engine are specified in~\cite{Willo21NatSupp}.}
\label{FWillomitzer:fig:NLoS_Schematics}
\end{figure}

\begin{figure}[b!]
\centering
\includegraphics[width=0.8\linewidth]{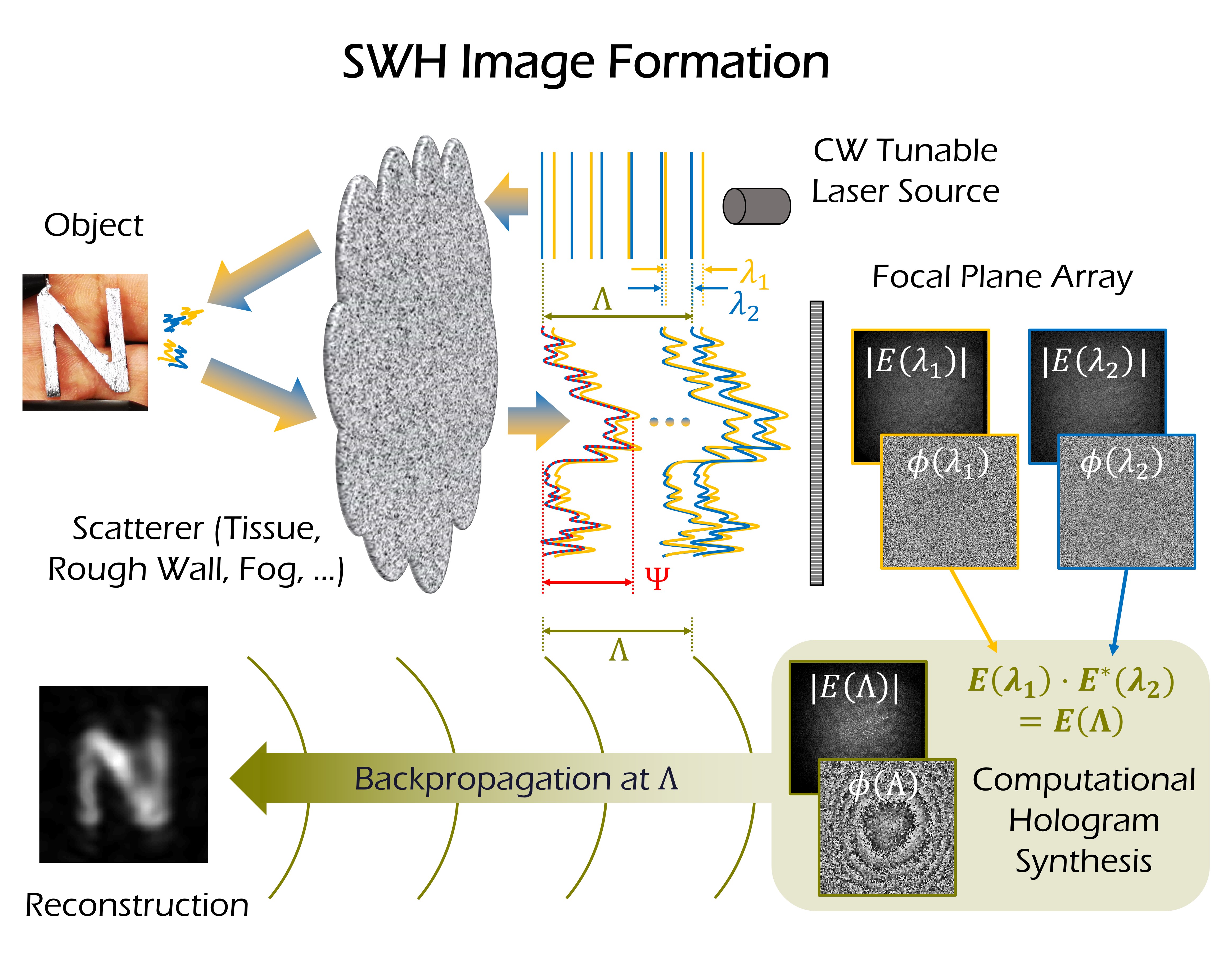}
\caption{Synthetic wavelength holography (SWH) image formation: The scene/object is illuminated with a continuous wave tunable laser at two slightly different wavelengths $\lambda_1$ and $\lambda_2$. Each field $E(\lambda_1)$, $E(\lambda_2)$ undergoes multiple scattering processes in
or at the scatterer (which could be a wall, tissue, fog,...) and the rough object surface. The introduced path length variation $\Psi$ leads to a 
randomization of the captured fields $E(\lambda_1)$ and $E(\lambda_2)$ and their respective phase maps $\phi(\lambda_1)$ and $\phi(\lambda_2)$. However, computational mixing of the speckled fields via Eq.~\ref{FWillomitzer:eq:Mixing} yields a complex-valued hologram $E(\Lambda)$ of the object at a synthetic wavelength $\Lambda$ (Eq.~\ref{FWillomitzer:eq:SWL}). The object is reconstructed by
backpropagating $E(\Lambda)$ with the synthetic wavelength $\Lambda$.}
\label{FWillomitzer:fig:SWH_Formation}
\end{figure}

It was discussed above that interferometry at the synthetic wavelength has a long tradition in optical metrology. This is also true for\index{Synthetic wavelength!holography} synthetic wavelength holography, which is used, e.g., for surface profiling of technical parts or industrial inspection (see~\cite{mann2008quantitative, fratz2021digital, fu2009dual, yamagiwa2018multicascade, hase2021multicascade, javidi2005three} for exemplary references). Similar to synthetic wavelength interferometry, the required synthetic phase maps can be captured in multiple ways, including methods that utilize spatial heterodyning, frequency heterodyning, or phase shifting~\cite{zhou2022review, kreis1997methods, kim2010principles}.

This section will focus on a specific novel application of synthetic wavelength holography which has been recently published in~\cite{Willo21Nat}: Imaging hidden objects around corners or through scattering media (which can be collectively referred to as "Non-Line-of-Sight" (NLoS) imaging)\index{Non-Line-of-Sight!imaging}\index{Non-Line-of-Sight!imaging!high resolution}\index{Non-Line-of-Sight!imaging!spectral}   at high resolution.
As before, \textit{spectral diversity}\index{Spectral diversity}, necessary to form the synthetic field $E(\Lambda)$, is utilized to mitigate the phase randomization in a speckle field. However, one of the new and important insights that have been demonstrated in~\cite{Willo21Nat, Willo19Arx, WilloCOSI19} is that the synthesis of a synthetic field $E(\Lambda)$ from two speckled optical fields is still possible if \textit{more than one} scattering process (e.g., in form of more than one optically rough surface) is involved. In this case, \textit{all} introduced microscopic path length variations contribute to the spectral decorrelation, and the synthetic wavelength has to be adjusted accordingly (see section~\ref{FWillomitzer:sec:Limits}).

Over the years many techniques have been proposed to image "hidden" objects, obscured from direct view~\cite{FREUND199049, WANG769, 2020NatRP...2..141Y, Ntz_2010, Dunsby_03, Hoshi_16, Yoo:s, Singh_14, Bertolotti_12, Mosk_12, Yaqoob_08, vellekoop_10, Xu_11, Doktofsky_20}. 
 Now, the problem is enjoying renewed attention. A solution can lead to potential applications in autonomous navigation, industrial inspection, planetary exploration, or early-warning systems for first-responders~\cite{Fac_NatPhys_19, Batarseh18, OtooleNat_18, MetzlerKeyhole, LindellWave, heide2014diffuse, LiuNat_19, Velten12, Faccio:19, katz14, katz2012looking, lindell20, gariepy_16, SAVA2019527}. Potential application scenarios are imaging through deep turbulence or fog, imaging through optically opaque barriers like skull, face identification around corners, and many more.

\subsection{ Imaging around Corners with Synthetic Wavelength Holography} 

\begin{figure}[b!]
\centering
\includegraphics[width=1\linewidth]{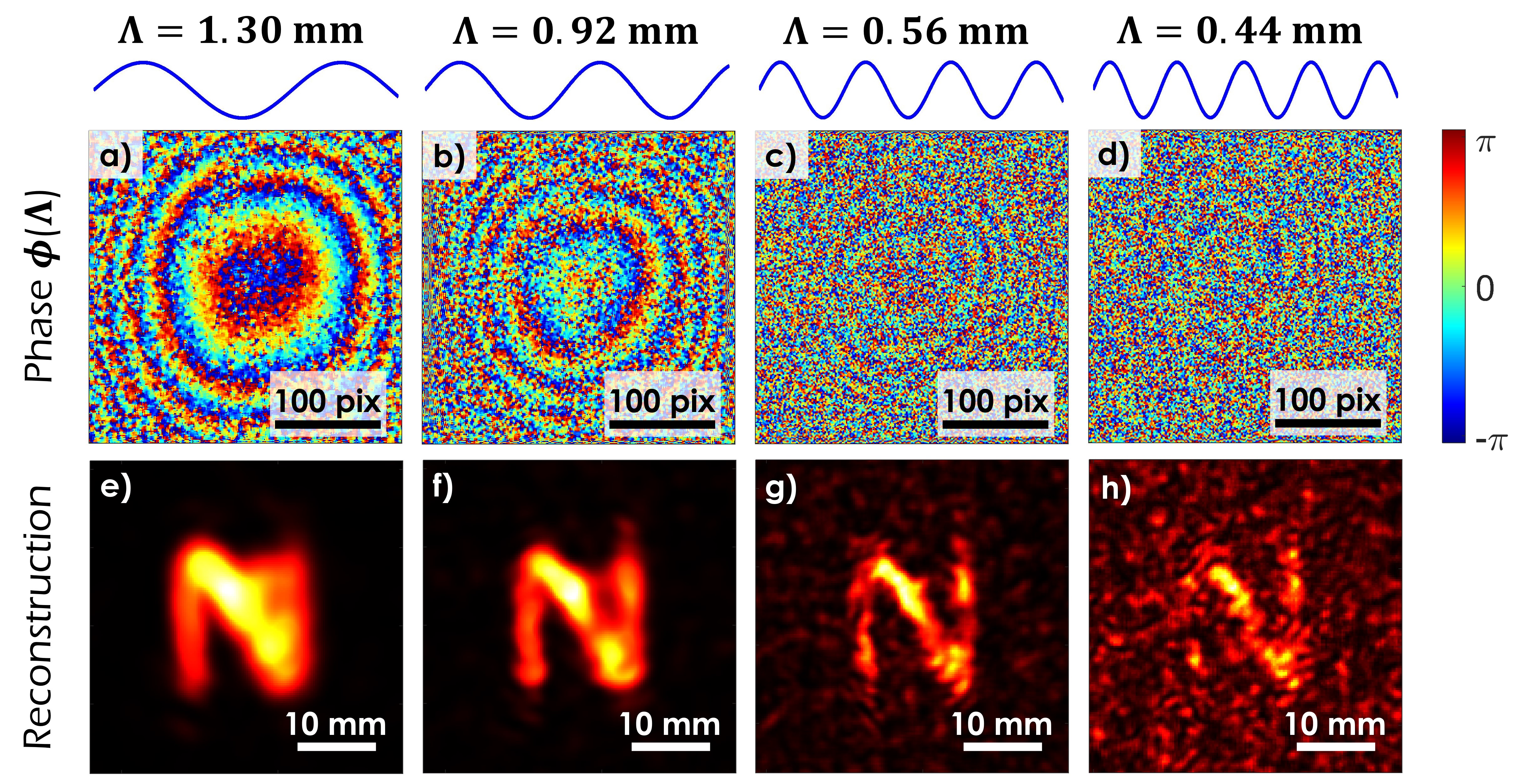}
\caption{Imaging around corners with synthetic wavelength holography (experiment). The character 'N' (dimensions $\sim 15mm \times 20mm $, photo shown in Fig.~\ref{FWillomitzer:fig:ObjScatt}a) is imaged around  the corner at four different synthetic wavelengths $\Lambda$. The setup schematic is shown in Fig.~\ref{FWillomitzer:fig:NLoS_Schematics}a. a-d) Phase maps of synthetic holograms captured at the VD surface. e-h) Respective reconstructions, displaying the squared magnitude of the backpropagated synthetic fields at the standoff distance of the character. The lateral resolution of the reconstructions increases with decreasing synthetic wavelength. However, the speckle-artifacts increase as well due to the decorrelation of the
two optical fields at $\lambda_1$ and $\lambda_2$.}
\label{FWillomitzer:fig:N_Result}
\end{figure}

The experimental setup\index{Synthetic wavelength!holography}\index{Imaging around corners} used to image hidden objects around corners is schematically depicted in Figure~\ref{FWillomitzer:fig:NLoS_Schematics}a. A reflective scatterer such as a wall is used to scatter light towards the hidden object and to intercept the back-scattered light. The respective portions of the wall are identified as "virtual source" (VS) and "virtual detector" (VD). The nomenclature "VS" and "VD" alludes to the fact that the method indeed synthesizes a virtual computational holographic camera (with source and detector) on the wall. The position of this virtual camera on the wall is chosen in a way that the hidden object resides \textit{in direct line of sight} of this virtual camera.
The FPA camera is focused on the VD portion of the wall, where the backscattered fields $E(\lambda_1)$ and $E(\lambda_2)$ are recorded. Eventually, the synthetic field $E(\Lambda)$ can be assembled, e.g., by mixing (see Eq.~\ref{FWillomitzer:eq:Mixing}). 
As $E(\Lambda)$ is captured at the VD position on the wall and the light waves are subject to an additional propagation between the hidden object and the wall, $E(\Lambda)$ represents now a \textit{hologram of the hidden object at the synthetic wavelength $\Lambda$}.
Again, if $\Lambda$ is chosen sufficiently large (see discussion in next section), the synthetic hologram $E(\Lambda)$ is not affected by speckle\index{Speckle} and shows a clear structure. A three-dimensional representation of the hidden object can be reconstructed by numerically backpropagating the assembled synthetic wavelength hologram $E(\Lambda)$ with the synthetic wavelength $\Lambda$. Figure~\ref{FWillomitzer:fig:SWH_Formation}  depicts the generalized image formation process.


Figures~\ref{FWillomitzer:fig:N_Result}a-d display the phase $\phi(\Lambda)$ of the computationally assembled synthetic wavelength hologram\index{Synthetic wavelength!measurements}, for a specific set of synthetic wavelengths $\Lambda$. For each measurement, it is possible to recover phase information, despite the pronounced multiple scattering at the object surface and wall. A 2D image of the final  reconstruction  is shown in Figs.~\ref{FWillomitzer:fig:N_Result}e-h. The images show the squared magnitude of the backpropagated synthetic holograms at the standoff distance of the object (character "N"). As expected, this is the backpropagation distance that produces the "sharpest" image of the character  for the respective synthetic wavelength. The experiments demonstrate the ability to recover an image of a small character `N' (dimensions $15mm \times 20mm$, see Fig.~\ref{FWillomitzer:fig:ObjScatt}a) despite being obscured from direct view.  Furthermore, the \textit{phase information} encapsulated in the synthetic hologram allows to evaluate the depth location of  the hidden object within the obscured volume. \\
As it can be seen in Figs.~\ref{FWillomitzer:fig:N_Result}e-h, the lateral resolution of the reconstruction improves with decreasing synthetic wavelength $\Lambda$. This behavior is in complete agreement with results from classical holography. It confirms again that the synthetic wave, although a computational construct, has distinct characteristics that it shares with a physical wave at the respective wavelength $\Lambda$.  The fundamental resolution limits of synthetic wavelength holography will be discussed in the next section.

\subsection{Imaging through Scattering Media with Synthetic Wavelength Holography} 

\begin{figure}[b!]
\centering
\includegraphics[width=1\linewidth]{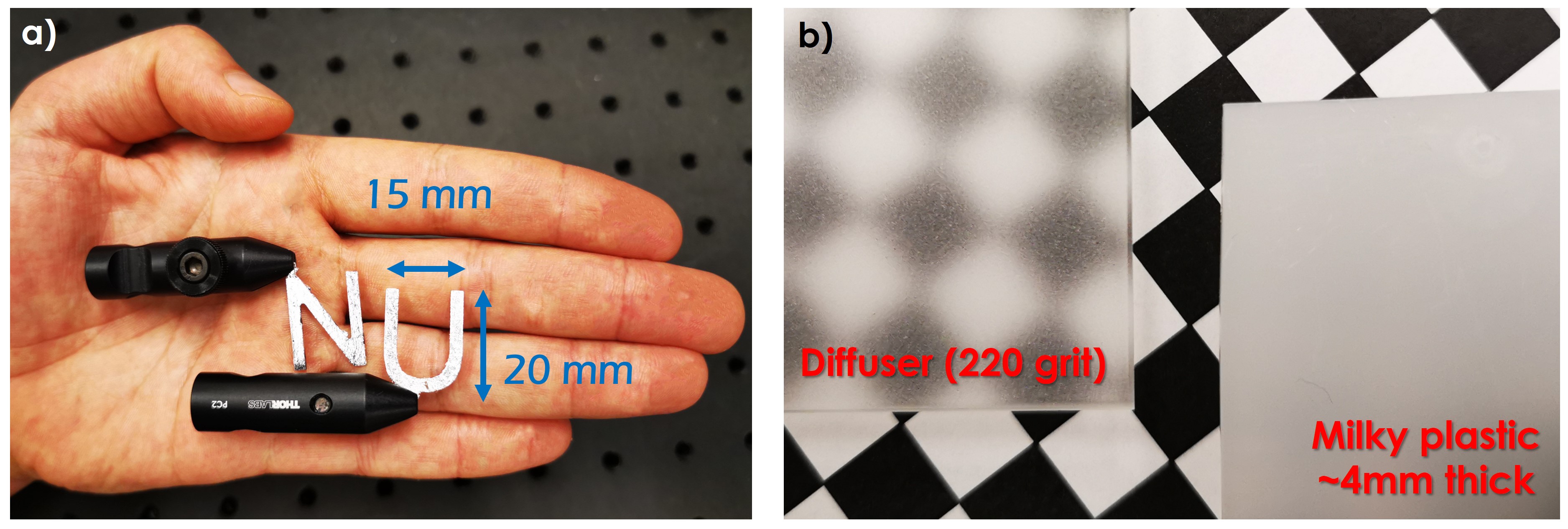}
\caption{a) Photo of the two objects (character 'N' and 'U') used for the shown synthetic wavelength holography experiments. Each character has the dimensions $ \sim 15mm \times 20mm$ (see hand for size comparison). b) Scatterers obscuring the 'U' character for the imaging through scattering media experiment shown in Fig.~\ref{FWillomitzer:fig:U_Result}:  A 220-grit ground glass diffuser and a milky white plastic plate of $\sim 4mm$ thickness, both placed $\sim 1cm$ over a printed  checker pattern to demonstrate the degradation in visibility.  }
\label{FWillomitzer:fig:ObjScatt}
\end{figure}

Besides\index{Synthetic wavelength!holography}\index{Imaging through scattering media} imaging hidden objects around corners, synthetic wavelength holography can also be  utilized to image hidden objects through a scattering medium, like skin, or fog. Although such a scenario is often considered the transmissive equivalent to the reflective "imaging around the corner" scattering  problem, the scattering can be much more severe in the transmissive case. Compared to the 2-3  distinct surface scattering processes for imaging around corners, light is commonly subject to much larger path length variations when penetrating through a volume scatterer. This makes imaging through strongly scattering media a much harder computational imaging problem.

\begin{figure}[b!]
\centering
\includegraphics[width=1\linewidth]{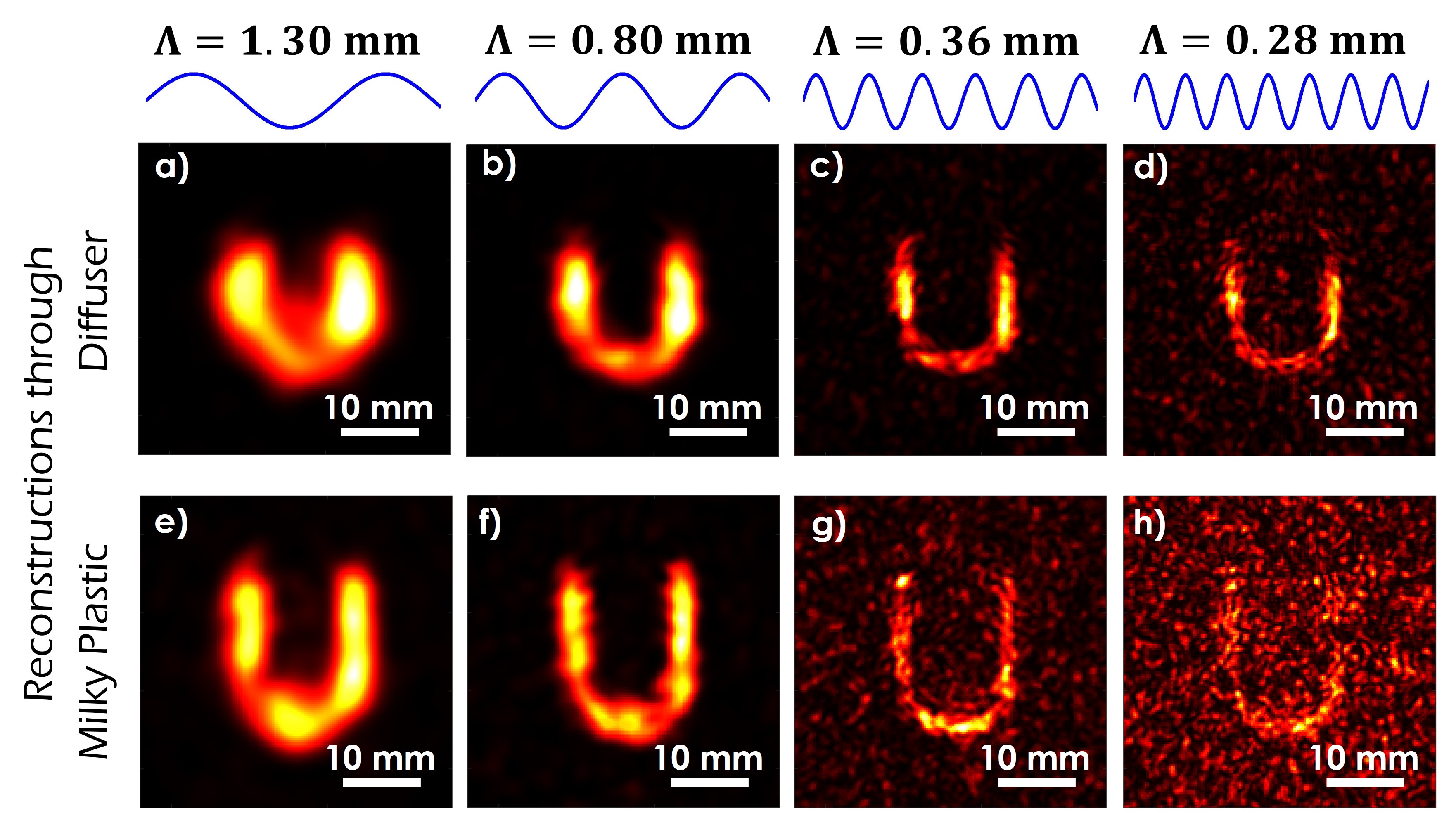}
\caption{Imaging through scatterers with synthetic wavelength holography (experiment). The imaged object is a  character 'U' (dimensions $\sim 15mm \times 20mm $, photo shown in Fig.~\ref{FWillomitzer:fig:ObjScatt}a). a-d) Reconstructions of measurements taken through a 220-grit ground glass diffuser (shown in Fig.~\ref{FWillomitzer:fig:ObjScatt}b) for different synthetic wavelengths~$\Lambda$. e-h) Reconstructions of measurements taken through a milky white plastic plate with $\sim 4mm$ thickness  (shown in Fig.~\ref{FWillomitzer:fig:ObjScatt}b) for different synthetic wavelength $\Lambda$.
 In analogy to Fig.~\ref{FWillomitzer:fig:N_Result}, smaller synthetic wavelengths $\Lambda$ deliver higher lateral resolution but increased speckle artifacts. The larger path length difference in the plastic plate leads to greater decorrelation.}
\label{FWillomitzer:fig:U_Result}
\end{figure}

The set of experiments shown in Fig.~\ref{FWillomitzer:fig:U_Result} demonstrates the versatility of the synthetic wavelength holography\index{Synthetic wavelength!measurements} principle by recovering holograms of objects hidden behind a scattering medium. The schematic setup is illustrated in Fig.~\ref{FWillomitzer:fig:NLoS_Schematics}b. The setup is equivalent to the setup depicted in Fig.~\ref{FWillomitzer:fig:NLoS_Schematics}a, but now adjusted to transmissive scattering. \\
In a first experiment, a small character `U' (dimensions $15mm \times 20mm$) is imaged through a 220 grit diffuser (Fig.~\ref{FWillomitzer:fig:ObjScatt}b left). The diffuser scatters light only at one of its surfaces. This means that this first experiment is indeed the transmissive equivalent to the "imaging around corners" scenario discussed above, as  the total scattering for this scenario can be again described as 2-3 distinct surface scattering processes.  Similar to the "imaging around corners" configuration, illumination and FPA camera are focused on the surface of the scatterer and form a VS and VD. Again, the  synthetic wavelength hologram $E(\Lambda)$ is captured at the VD surface,  and the object is reconstructed via numerical backpropagation at the synthetic wavelength $\Lambda$.
 The holographic reconstructions of the  character `U' are shown in Fig.~\ref{FWillomitzer:fig:U_Result}a-d. Again, the lateral resolution of the reconstruction increases with decreasing synthetic wavelength $\Lambda$. \\
In a second experiment, the ground glass diffuser in the imaging path is swapped with a $4mm$ thick milky plastic plate (Fig.~\ref{FWillomitzer:fig:ObjScatt}b right). The plate can be considered a strong volume scatterer, as the light undergoes several scattering processes during its transmission. This becomes apparent in Fig.~\ref{FWillomitzer:fig:ObjScatt}b by comparing the degraded visibility of a checkerboard pattern  which is viewed through the plastic plate and the 220 grit diffuser for comparison. However, despite pronounced  scattering in the plastic plate, the character `U' can be reconstructed for synthetic wavelengths exceeding $360\mu m$, as shown in Figs.~\ref{FWillomitzer:fig:U_Result}e-h. This confirms  the ability to recover image information at visibility levels far below the perceptual threshold. A comparison of the reconstructions for the plastic plate and the diffuser reveals only a marginal change in the smallest achievable synthetic wavelength. This observation can be explained by the fact that the visibility of ballistic light paths decays exponentially with the propagation distance through a scattering volume (in accordance with Beer's law~\cite{Beer}), whereas the lateral resolution for synthetic wavelength holography is  linearly related to the choice of $\Lambda$, as discussed in the next section.

\subsection{Discussion and Comparison with the State of the Art}

As mentioned before, the problem of "Non-Line-of-Sight" imaging, which is here collectively referred to as the task of imaging around corners and imaging through scattering media, has recently enjoyed renewed attention. Besides the introduced synthetic wavelength holography technique, existing active methods are either based on ToF imaging ("transient techniques," see also earlier chapters of this book) or exploit spatial correlations in the scattered optical fields, i.e., the so-called spatial (or angular) "memory effect"~\cite{Freund_88, GoodmanSpeck}. \\
Recent work in the area of ToF-based techniques using transients\index{Transient} has  demonstrated results with $cm$-scale lateral resolution over a $\sim 1m\times 1m\times 1m$ working volume, and in select cases providing near real-time reconstructions. However, many approaches rely on raster scanning large areas on VS and/or VD whose dimensions are comparable to the obscured volume~\cite{OtooleNat_18, LiuNat_19, Velten12, Faccio:19, LindellWave, heide2014diffuse, nam2021low}. \\
 Spatial correlation-based\index{Correlography} techniques allow for the highest lateral resolution of object reconstructions (${< 100\mu m}$ at $1m$ standoff). Moreover, the probing area on the intermediary VS/VD surface can be less than a  few $cm$. These benefits, however, come at the expense of a highly restricted angular field of view ($<2^\circ$), as determined by the angular decorrelation of scattered light.
This angular memory effect does not only limit the field of view but also the maximal possible size of the measured object, which is not allowed to exceed the respective working volume~\cite{Edrei:16, Singh_14, katz14, PrasannaSPIE, Aparna_Corr, Balaji:17, metzler2020deep}.

\begin{figure}[h!]
\centering
\includegraphics[width=0.6\linewidth]{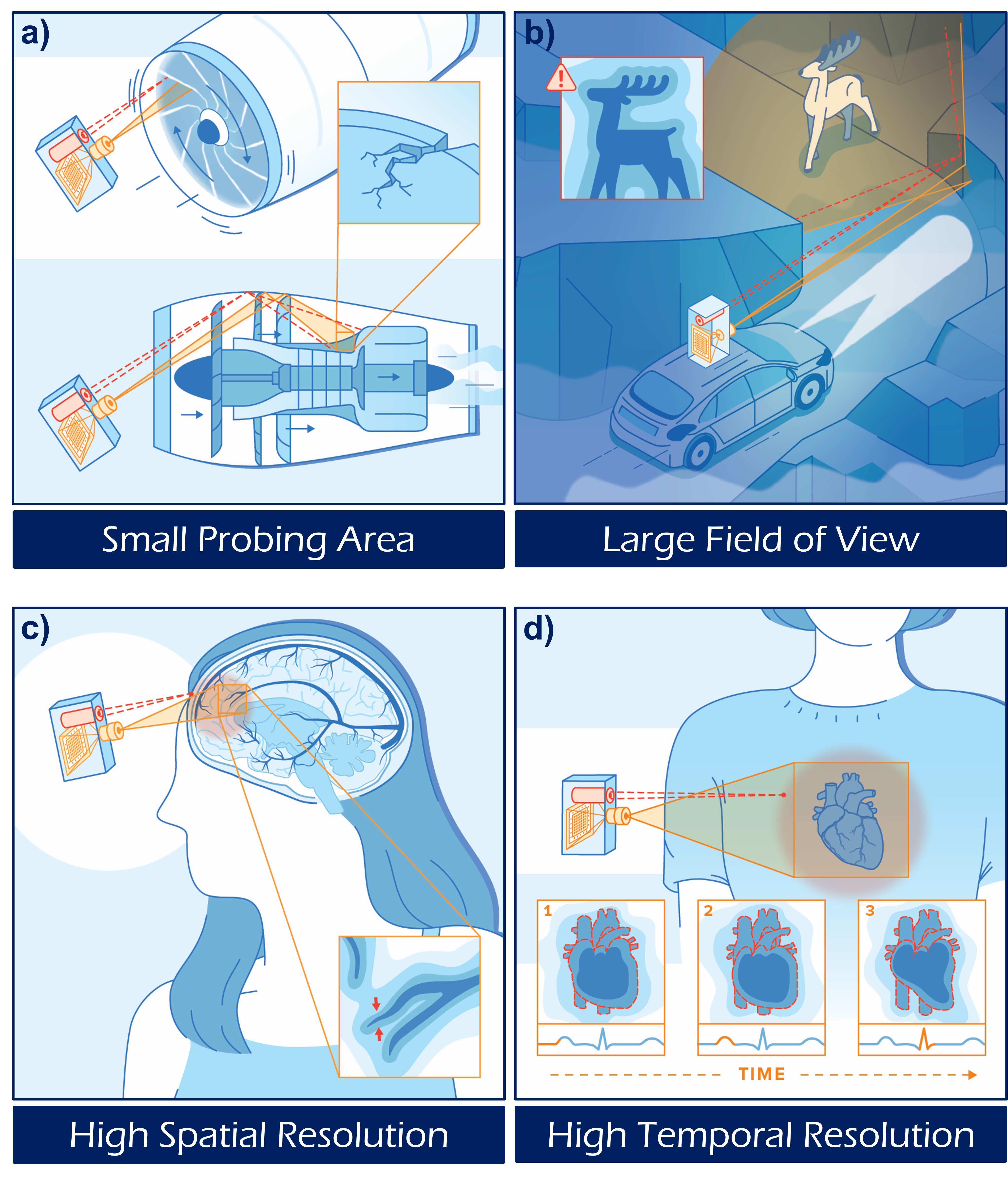}
\caption{Key attributes of synthetic wavelength holography (SWH) and potential future applications. The SWH approach combines four key attributes, each shown  in a potential
future NLoS application example: in each scenario, a scattering surface or medium is used to indirectly illuminate, and intercept light scattered by the
hidden object. a)  A small probing area allows for the inspection of defects in tightly confined spaces, e.g., in running aircraft engines. b) A wide angular field of view allows to measure/detect hidden objects without previous knowledge of their position as, e.g., important when navigating in degraded visual environments. c) High
spatial resolution allows for the measurement of small structures, such as non-invasive imaging of brain vessels through the skull. d) High temporal
resolution allows to image objects in motion, e.g., to potentially discern cardiac arrhythmia through the chest. Synthetic wavelength holography combines all of these four attributes in one single approach to NLoS imaging. }
\label{FWillomitzer:fig:Attributes}
\end{figure}

The wide disparity in achievable field of view and resolution of other NLoS imaging techniques can limit their usability. In contrast, the introduced synthetic wavelength holography technique allows for a combination of capabilities that is, to the author's knowledge, currently unmatched by the state of the art~\cite{Willo21Nat}. The respective attributes (also depicted in Fig.~\ref{FWillomitzer:fig:Attributes}) are:

\begin{itemize}

\item 

\textbf{Small Probing Area:}
Many transient-based NLoS schemes require probing areas (VD or VS sizes) with dimensions around $\sim 1m\times 1m$ which limits their ability to detect hidden objects in confined spaces. 
Synthetic wavelength holography provides the ability to image obscured objects by simultaneously illuminating and observing a small area ($58mm\times58mm$ for the shown experiments).

\item 

\textbf{Wide Angular Field of View:}
Angular memory effect-based approaches are limited to  highly restricted fields of view ($< 2^\circ$ for drywall). As a holographic method, synthetic wavelength holography provides the ability  to recover obscured objects over a nearly hemispherical field of view that far exceeds the limited angular extent of the memory effect.

\item

\textbf{High Spatial Resolution:}
Transient or ToF camera-based  approaches generally produce rather low spatial resolutions ($\sim cm$), due to the long modulation wavelengths. Synthetic wavelength holography provides the ability to resolve small features on obscured objects (up to ${<1mm}$ in the shown experiments). 

\item

\textbf{High Temporal Resolution:}
Many transient-based  approaches rely on point-wise raster-scanning. Synthetic wavelength holography is able to recover full field holograms of the obscured object  using off-the-shelf FPA technology. The synthetic wavelength holography principle even allows for single-shot acquisition. 

\end{itemize}

So far for the specific attributes of synthetic wavelength holography. The following section will discuss an important topic that has been largely omitted so far: \textit{Fundamental performance limits of synthetic wavelength imaging} in general.
It already became clear from the preceding explanations that the size of the synthetic wavelength $\Lambda$, as well as the severity of scattering imparted by the rough surfaces and the scattering media, plays an important role in this discussion. It will be explained how these parameters and other system parameters like the VD size influence the resolution of the respective methods, and the related trade-off space will be explored.

\section{Fundamental Performance Limits of Synthetic Wavelength Imaging} 
\label{FWillomitzer:sec:Limits}


It\index{Limits!physical}\index{Limits!fundamental performance} has been mentioned at the beginning of this chapter that the depth precision of ToF-based methods like Interferometry or CWAM ToF imaging is directly proportional to the modulation wavelength (Eq.~\ref{FWillomitzer:eq:PrecMod}). The discussion has been further backed up by the notion that a generalized wave concept can be indeed used to discuss general trends of the performance (such as the trend of increasing precision with decreasing modulation wavelength)~\cite{reza2019phasor, gupta2015phasor, Willo21Nat}. For this general discussion, the \textit{origin} of the wave (complex electromagnetic field, optical beat, amplitude modulation, etc.) became of secondary importance. The aim for higher depth precision and/or lateral resolution was one of the main motivation points to switch from a meter-sized amplitude modulated wave (as used in CWAM ToF cameras) to a synthetic wave with a much smaller wavelength $\Lambda$. However, it also became clear that $\Lambda$ cannot become indefinitely small: As seen in single-wavelength interferometry, measurements at a very small (i.e., optical) wavelength are subject to phase randomization in a speckle field. The inevitable question:

\noindent \textit{How small can the synthetic wavelength $\Lambda$ become, and what are the related limits?}

A\index{Synthetic wavelength!smallest possible} well-known model to describe the severity of scatter and the related formation of a speckle field for \textit{optical waves} is the\index{Rayleigh quarter wavelength criterion} so-called "Rayleigh quarter wavelength criterion"~\cite{Rayleigh}.  According to this criterion, an optical field that is reflected off a rough surface or surpasses a scattering medium forms speckle if the maximal path length variation $\Psi$ introduced by the scatterer and the geometry exceeds $1/4$ of the wavelength $\lambda$ within one object-sided diffraction disc\footnote{The size of the object-sided diffraction is the size of the  image-sided diffraction disc divided by the magnification of the optical imaging system. It can be seen as the projection of an image-sided diffraction disc onto the object surface}.

\begin{equation}
   \Psi < \frac{\lambda}{4} ~~~~ \textrm{for speckle-free imaging}
\end{equation}
This criterion is intuitively understandable: A light field  that exhibits random path length variations of maximal $\Psi= \lambda/4$ while propagating from the source to the object, and another $\Psi= \lambda/4$ for propagating from the object back to the detector is subject to  spatially varying random phase shifts of maximal $\pi$. A phase shift of $\pi$ is just large enough for fully destructive interference, i.e.,  an\index{Speckle}\index{Speckle!formation} interference pattern at full contrast~\cite{GH04Enc, hausler2011limitations}. \\
It should be noted for the sake of completeness, that the Rayleigh quarter wavelength criterion is also frequently used vice versa to define whether a surface\index{Surface roughness}\index{Surface roughness!definition} is optically rough or not. In this case, a surface can be defined as "rough" if the path length variations within one object-sided diffraction disc exceed $\lambda/4$. This leads to the interesting fact that the roughness definition not only depends on the surface itself but also on the geometry (tilt) and the resolving power of the optical imaging apparatus. The direct follow-up question is \textit{which statistical definition of "roughness" should be used}. Strictly spoken, the first destructive interferences (dark speckles) should appear if the peak-to-valley surface roughness $R_p$ exceeds $\lambda/4$. However, as the microscopic height values of most surfaces are normal distributed, this will happen very infrequently, which is the reason why most definitions use the RMS surface roughness $\sigma_h$ instead. The RMS surface roughness $\sigma_h$ will also be used here for further definitions.

So far for imaging at the optical wavelength $\lambda$. Although some analogies between the synthetic wavelength and the optical wavelength have been drawn already, related  limit considerations for the synthetic wavelength $\Lambda$ seem less intuitive. This is particularly true for such realizations of  synthetic wavelength imagers that acquire both optical fields  $E(\lambda_1)$ and $E(\lambda_2)$ in a sequential fashion (e.g., dual-wavelength heterodyne interferometers). In this case, the two optical fields did never physically beat together, and the synthetic wave becomes a purely computational construct that is generated in the computer via post-processing (Eq.~\ref{FWillomitzer:eq:Mixing}). Nevertheless, it has been shown in~\cite{Willo21NatSupp} that the Rayleigh quarter wavelength criterion \textit{can  still be applied} to measurements at the synthetic wavelength.
For artifact-free imaging, this means that the smallest possible synthetic wavelength $\Lambda$ can be estimated via

\begin{equation}
   \Psi < \frac{\Lambda}{4}  ~~~~ \Rightarrow  ~~~~  \Lambda > 4 \cdot \Psi~~.
   \label{FWillomitzer:eq:RQWCSW}
\end{equation}

Reconstructions obtained from  measurements at  $\Lambda$ close to or smaller than $4 \Psi$ start to exhibit speckle-like artifacts (see, e.g., Fig.~\ref{FWillomitzer:fig:N_Result}h or Fig.~\ref{FWillomitzer:fig:U_Result}g,h) which become more severe with decreasing $\Lambda$. Indeed, these artifacts can be interpreted as \textit{"synthetic speckle!"}

This is a \textbf{\textit{remarkable analogy}} which requires further explanation: It has been discussed that an artifact-free synthetic phase map $\phi(\Lambda)$ can be obtained if the two optical (speckle) fields $E(\lambda_1)$ and $E(\lambda_2)$ are sufficiently correlated.  For the  synthetic wavelength imaging methods introduced here, the two object beams originate from the same point (same fiber tip), and the respective speckle patterns change their appearance with different wavelengths. Hence, this correlation can be understood as a \textit{spectral correlation}. The related effect of spectral decorrelation shows a strong analogy to the "memory effect" for angular decorrelation~\cite{Freund_88, GoodmanSpeck}. In fact, it can be understood as \textit{"spectral memory effect"}~\cite{GoodmanSpeck, Willo21Nat}. The transition from correlated to uncorrelated fields is, of course, a fluent and not binary process, meaning that different criteria exist regarding whether two fields can be treated as "correlated" or not~\cite{GoodmanSpeck}. Applying similar decorrelation criteria used to define the isoplanatic angle for the angular memory effect leads to a maximal wavelength separation $|\lambda_1 - \lambda_2|$, which is dependent on the maximal path length variation $\Psi$ and the "starting" wavelength $\lambda_1$~\cite{GoodmanSpeck}. Combined with Eq.~\ref{FWillomitzer:eq:SWL}, this results in Eq.~\ref{FWillomitzer:eq:RQWCSW}. The concrete calculations and further details can be found in~\cite{Willo21NatSupp}.

\textit{What does the limitation of Eq.~\ref{FWillomitzer:eq:RQWCSW} mean for synthetic wavelength interferometry and synthetic wavelength holography?} In both cases, the maximal path length variation $\Psi$ introduced by the scatterer restricts the smallest possible synthetic wavelength $\Lambda$\index{Synthetic wavelength!holography!fundamental limits}\index{Synthetic wavelength!interferometry!fundamental limits}.

\noindent \textbf{For synthetic wavelength interferometry} on pure surface scatterers,  the path length variations are introduced by the surface roughness and the geometry. A beam that hits the rough surface from an oblique angle is subject to different path length variations than a beam at normal incidence. In the limit case of scanning the surface at a single point with beam diameter $\lambda_1 \approx \lambda_2$ and normal incidence, the path length variation converges into the RMS surface roughness ($\Psi_{Int} \rightarrow \sigma_h$)~\cite{Willo21NatSupp}.  This results in the criterion

\begin{equation}
    \Lambda > 4 \cdot \Psi_{Int} > 4 \sigma_h~~.
\end{equation}

\noindent \textbf{For imaging around the corner with synthetic wavelength holography}, the coherent field exhibits \textit{at least} two scattering processes on a rough surface (the wall). Assuming again no subsurface scattering, normal incidence, and a VS/VD diameter of $\lambda_1 \approx \lambda_2$, the path length variation converges to  $\Psi_{Hol1} \rightarrow 2 \cdot \sigma_h$ ~\cite{Willo21NatSupp}, which means that

\begin{equation}
    \Lambda > 4 \cdot \Psi_{Hol1} > 8 \cdot \sigma_h~~.
\end{equation}

\noindent  \textbf{For imaging through scattering media with synthetic wavelength holography}, it is assumed that the light traverses two times (round trip) through a volume scattering medium with thickness $L$ and transport mean free path $l^*$. In the described limit case, the path length variation converges to  $\Psi_{Hol2} \rightarrow 2 \cdot L^2/l^*$ ~\cite{Willo21NatSupp}, leading to:

\begin{equation}
    \Lambda > 4 \cdot \Psi_{Hol2} > 8 \cdot \frac{L^2}{l^*}
\end{equation}

Having established these bounds, it can finally be discussed how the size of the smallest possible synthetic wavelength is tied to the resolution of the respective method. It is known from metrology literature that well-calibrated single wavelength interferometers can reach impressive depth precisions of $\delta z = \lambda /100$ (or even better) on specular surfaces. A depth error of, e.g., $\lambda /100$ means that the optical phase can be determined with a precision of $\delta \phi = 2\pi / 100$. However, caution is advised in extrapolating this result to synthetic wavelength interferometry on rough surfaces and assuming that precisions of $\delta z = \Lambda /100$ can easily be reached as well! For every point in the camera image,  waves are accumulated from a small surface area on the object (the object-sided diffraction disc). Per previous definition, the surface topography of rough surfaces varies significantly within this object-sided diffraction disc. The consequence is a large uncertainty in the measured optical phase, which translates to the synthetic phase. This is the reason why the precision of multi-wavelength interferometry on rough surfaces is ultimately bound by the surface roughness~\cite{Dresel:92, hausler1994range, haeusler1993coherence, ettl1998roughness} and can typically not reach the very high phase precision $\delta \phi$ of  single-wavelength interferometry on smooth surfaces. It is referred to~\cite{GH22LAM} for further information about this topic. Reference~\cite{Li21} contains a precision evaluation of the introduced synthetic wavelength interferometer for macroscopic objects with rough surfaces. 

For the introduced technique of synthetic wavelength holography to image around corners or through scattering media, the lateral resolution and longitudinal  localization uncertainty are influenced by the geometry of the setup and the size of the synthetic wavelength $\Lambda$ . To reconstruct an image of the hidden object, the field captured at the VD is back-propagated in the hidden volume. In analogy to light focusing through a lens, we can approximate a numerical aperture by the radius of the virtual detector $D_{VD} / 2$ divided by the standoff distance $z$. Analog to classical optics, the lateral resolution $\delta x$ can then be approximated as~\cite{Willo21Nat}

\begin{equation}
    \delta x \approx \frac{\Lambda z}{D_{VD}} ~~.
    \label{FWillomitzer:eq:LatRes}
\end{equation}

\begin{figure}[t]
\centering
\includegraphics[width=0.9\linewidth]{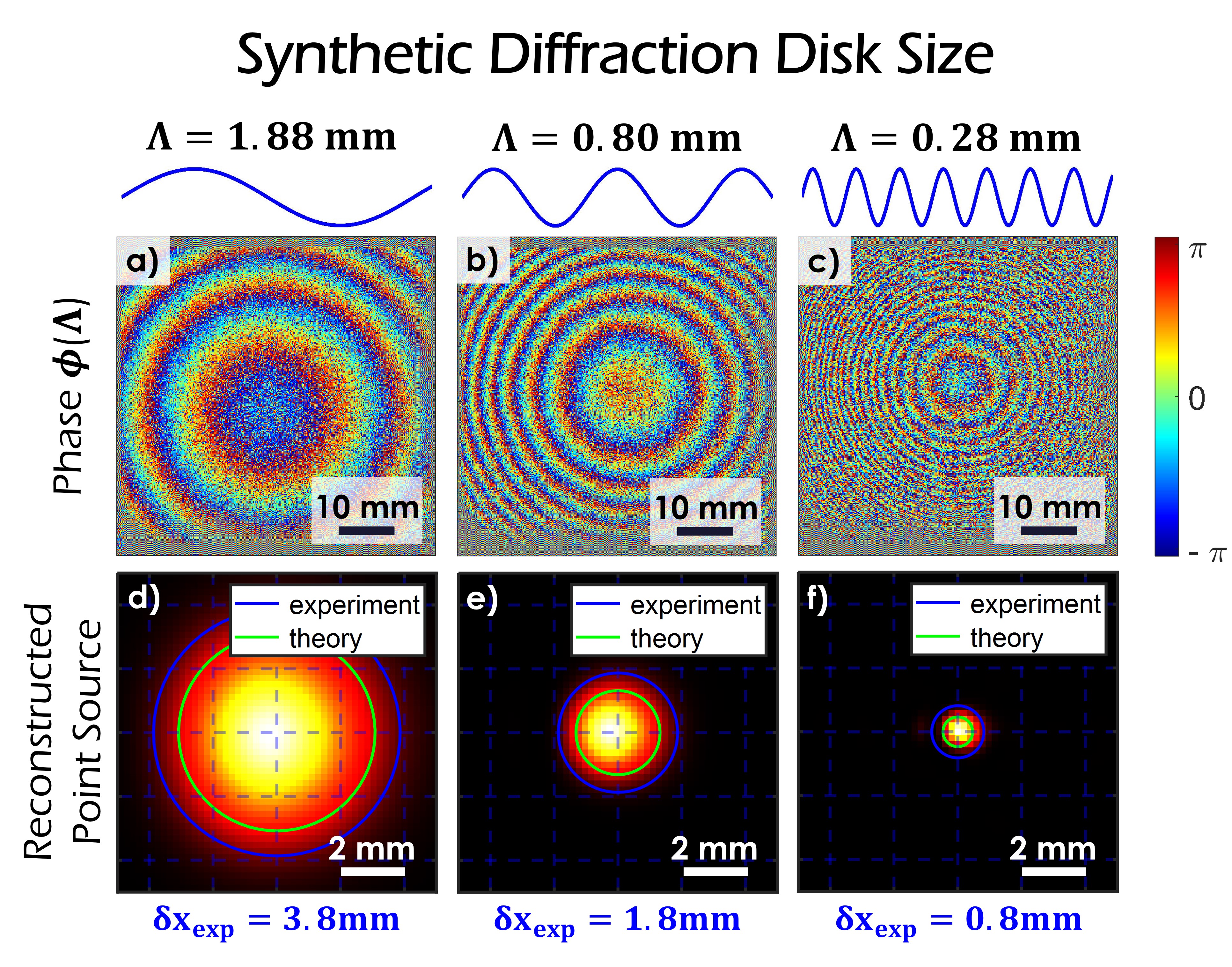}
\caption{Measurement of the "synthetic diffraction disc" via reconstruction of an obscured point source for three different synthetic wavelengths (experiment). a-c) Phase of the
synthetic holograms captured at the VD surface. The diameter of the VD is $D_{VD} = 58mm$ and the standoff distance of the point source (fiber tip) is $z = 95mm$. d-e) Reconstruction of the "synthetic diffraction disc." In agreement with classical optics (Eq.~\ref{FWillomitzer:eq:LatRes}), the disc size varies linearly with the
wavelength (in this case, the synthetic wavelength). The experimental value is close to the theoretical expectation. For $\Lambda = 0.28mm$ (f), the point source is reconstructed with sub-mm lateral resolution. }
\label{FWillomitzer:fig:SynthDiffDisc}
\end{figure}

Figure~\ref{FWillomitzer:fig:SynthDiffDisc} displays experimental results to evaluate the lateral  resolution $\delta x$ around the corner. A surface patch on the rough wall (the VD) was illuminated by a point-like light source (a fiber tip), which is located in the hidden volume. The patch has a diameter of $D_{VD} = 58mm$ and the fiber tip is located at a standoff distance of $z = 95mm$. After illuminating the VD with two wavelengths $\lambda_1$ and $\lambda_2$ and forming the synthetic hologram, the point-like light source was reconstructed by numerical back-propagation at the synthetic wavelength~$\Lambda$. It can be seen that the diameter of the "synthetic diffraction disc"\index{Synthetic wavelength!diffraction disc}\index{Synthetic diffraction disc} decreases with decreasing synthetic wavelength and that the general trend closely follows the theoretical expectation from Eq.~\ref{FWillomitzer:eq:LatRes}. For the measurement  at $\Lambda = 0.28mm$, the experimentally evaluated lateral resolution around the corner is \textit{below} $1mm$ (Fig.~\ref{FWillomitzer:fig:SynthDiffDisc}c,f). \\
Besides the discussed lateral resolution  $\delta x $, the longitudinal localisation uncertainty $\delta z$ behaves also as expected from classical optics and goes proportional to $ \delta z \sim \Lambda z^2 / D_{VD}^2$. The exact value  is also strongly dependent on the noise properties of the system. This is, of course, the case for all of the introduced precision and resolution measures, although it was not discussed in detail in this book chapter. The reader is referred to~\cite{Willo21Nat, Li21} and future publications of the author's research group. 
Similar to what has been discussed for synthetic wavelength interferometry, the longitudinal localization uncertainty of synthetic wavelength holography can be improved by the use of \textit{multiple} synthetic wavelengths. A related experiment ("synthetic pulse holography")\index{Synthetic wavelength!pulse holography}\index{Synthetic pulse holography} is described in~\cite{Willo21Nat}: The hidden scene is interrogated at multiple synthetic wavelengths, and the respective back-propagated (complex) fields are superpositioned coherently in the computer. As this procedure mimics the computational synthesis of a "synthetic pulse train," the longitudinal localization uncertainty can be significantly improved. It should also be noted that this technique shows striking similarities to optical coherence tomography (OCT) or white light interferometry (WLI), which will be further investigated in the future. Again, the reader is referred to~\cite{Willo21Nat, Willo19Arx} for additional information and further references.

\section{Conclusion and future directions}

This book chapter has discussed how spectral correlations in scattered light fields can be utilized for high-precision ToF sensing. It was shown that the introduced synthetic wavelength imaging techniques are able to extract phase information from optical speckle fields, which are subject to heavy scattering. The related techniques of synthetic wavelength interferometry and synthetic wavelength holography have long been  known in optical metrology, and have been used, e.g., for industrial inspection and surface testing. This book chapter has outlined how to apply these methods to novel problems in computational imaging and computer vision, such as high-precision ToF imaging for AR/VR and medical applications, or the problem of imaging objects around corners and through scattering media. It has been demonstrated that the introduced techniques can achieve very high depth precision and lateral resolution (in many cases much higher than established methods), and show a unique combination of other valuable attributes, such as a wide field of view, or high temporal resolution. \\
However, the discussions in the last section revealed that the introduced synthetic wavelength imaging approaches are not without limitations. The good news: Physical limitations often come in the shape of uncertainty products! This makes it possible to optimize a technique towards a specific quantity (e.g., speed or resolution) by trading in information less critical for the targeted application. These limitations and their tradeoff spaces will be further explored and exploited by the author's research group in the future.

\noindent
\textbf{Acknowledgement}: Research is always a team effort! Hence, it should be emphasized that the presented results and insights are the product of a continuous process that involves many students and colleagues and have not been produced and derived by the author alone.
Amongst others, these students and colleagues are Muralidhar M. Balaji,  Fengqiang Li, Marc P. Christensen, Manuel Ballester, and Heming Wang. From the involved colleagues, the author would particularly thank Oliver Cossairt, Prasanna Rangarajan, and Gerd H{\"a}usler for proofreading this book chapter, for their valuable comments, and for the always exciting discussions.

\bibliographystyle{unsrt}
\bibliography{references}

\begin{thebibliography}{100}

\bibitem{Srinivasan84}
Venugopal Srinivasan, Hsin-Chu Liu, and Maurice Halioua.
\newblock Automated phase-measuring profilometry of 3-d diffuse objects.
\newblock {\em Applied optics}, 23(18):3105--3108, 1984.

\bibitem{Schaffer10}
Martin Schaffer, Marcus Grosse, and Richard Kowarschik.
\newblock {High-speed pattern projection for three-dimensional shape
  measurement using laser speckles}, Appl. Opt. 49(18), 3622-3629 (2010).

\bibitem{willomitzer20103d}
Florian Willomitzer, Zheng Yang, Oliver Arold, Svenja Ettl, and Gerd
  H{\"a}usler.
\newblock 3d face scanning with "flying triangulation.".
\newblock {\em DGaO Proc}, 111:18, 2010.

\bibitem{Takeda83}
Mitsuo Takeda and Kazuhiro Mutoh.
\newblock Fourier transform profilometry for the automatic measurement of 3-d
  object shapes.
\newblock {\em Appl. Opt.}, 22(24):3977--3982, Dec 1983.

\bibitem{schoenberger16}
Johannes~Lutz Sch\"{o}nberger and Jan-Michael Frahm.
\newblock {Structure-from-Motion Revisited}.
\newblock In {\em Conference on Computer Vision and Pattern Recognition
  (CVPR)}, 2016.

\bibitem{arold2014hand}
Oliver Arold, Svenja Ettl, Florian Willomitzer, and Gerd H{\"a}usler.
\newblock Hand-guided 3d surface acquisition by combining simple light
  sectioning with real-time algorithms.
\newblock {\em arXiv preprint arXiv:1401.1946}, 2014.

\bibitem{WilloOE_17}
Florian Willomitzer and Gerd H\"{a}usler.
\newblock Single-shot 3d motion picture camera with a dense point cloud.
\newblock {\em Opt. Express}, 25(19):23451--23464, Sep 2017.

\bibitem{ettl2013improved}
Svenja Ettl, Stefan Rampp, Sarah Fouladi-Movahed, Sarang~S Dalal, Florian
  Willomitzer, Oliver Arold, Hermann Stefan, and Gerd H{\"a}usler.
\newblock Improved eeg source localization employing 3d sensing by "flying
  triangulation".
\newblock In {\em Videometrics, Range Imaging, and Applications XII; and
  Automated Visual Inspection}, volume 8791, pages 194--200. SPIE, 2013.

\bibitem{Woodham80}
Robert~J Woodham.
\newblock Photometric method for determining surface orientation from multiple
  images.
\newblock {\em Optical engineering}, 19(1):139--144, 1980.

\bibitem{Horn90}
Berthold~KP Horn.
\newblock Height and gradient from shading.
\newblock {\em International journal of computer vision}, 5(1):37--75, 1990.

\bibitem{Willo20PMD}
Florian Willomitzer, Chia-Kai Yeh, Vikas Gupta, William Spies, Florian
  Schiffers, Aggelos Katsaggelos, Marc Walton, and Oliver Cossairt.
\newblock Hand-guided qualitative deflectometry with a mobile device.
\newblock {\em Opt. Express}, 28(7):9027--9038, Mar 2020.

\bibitem{Knauer04}
Markus~C. Knauer, J\"urgen Kaminski, and Gerd H\"ausler.
\newblock Phase measuring deflectometry: a new approach to measure specular
  free-form surfaces.
\newblock In {\em Proc.SPIE}, volume 5457, pages 5457 -- 5457 -- 11, 2004.

\bibitem{faber12}
Christian Faber, Evelyn Olesch, Roman Krobot, and Gerd H\"ausler.
\newblock Deflectometry challenges interferometry: the competition gets
  tougher!
\newblock In {\em Proc.SPIE}, volume 8493, pages 8493 -- 8493 -- 15, 2012.

\bibitem{Wang21}
Jiazhang Wang, Bingjie Xu, Tianfu Wang, Wung~Jae Lee, Marc Walton, Nathan
  Matsuda, Oliver Cossairt, and Florian Willomitzer.
\newblock Vr eye-tracking using deflectometry.
\newblock In {\em Computational Optical Sensing and Imaging}, pages CF2E--3.
  Optical Society of America, 2021.

\bibitem{Huang2018}
Lei Huang, Mourad Idir, Chao Zuo, and Anand Asundi.
\newblock Review of phase measuring deflectometry.
\newblock {\em Optics and Lasers in Engineering}, 107:247 -- 257, 2018.

\bibitem{GH22LAM}
Gerd H{\"a}usler and Florian Willomitzer.
\newblock Reflections about the holographic and non-holographic acquisition of
  surface topography: where are the limits?
\newblock {\em Light: Advanced Manufacturing}, 3(2):1--10, 2022.

\bibitem{collis1970lidar}
RTH Collis.
\newblock Lidar.
\newblock {\em Appl. Opt.}, 9(8):1782--1788, 1970.

\bibitem{weitkamp2006lidar}
Claus Weitkamp.
\newblock {\em Lidar: range-resolved optical remote sensing of the atmosphere},
  volume 102.
\newblock Springer Science \& Business, 2006.

\bibitem{schwarte1997new}
Rudolf Schwarte, Zhanping Xu, Horst-Guenther Heinol, Joachim Olk, Ruediger
  Klein, Bernd Buxbaum, Helmut Fischer, and Juergen Schulte.
\newblock New electro-optical mixing and correlating sensor: facilities and
  applications of the photonic mixer device (pmd).
\newblock In {\em Proc. SPIE}, volume 3100, pages 245--254, 1997.

\bibitem{lange2001solid}
Robert Lange and Peter Seitz.
\newblock Solid-state time-of-flight range camera.
\newblock {\em IEEE J. Quantum Electron.}, 37(3):390--397, 2001.

\bibitem{foix2011lock}
Sergi Foix, Guillem Alenya, and Carme Torras.
\newblock Lock-in time-of-flight (tof) cameras: a survey.
\newblock {\em IEEE Sens. J.}, 11(9), 2011.

\bibitem{Li21}
Fengqiang Li, Florian Willomitzer, Muralidhar~Madabhushi Balaji, Prasanna
  Rangarajan, and Oliver Cossairt.
\newblock Exploiting wavelength diversity for high resolution time-of-flight 3d
  imaging.
\newblock {\em IEEE Transactions on Pattern Analysis and Machine Intelligence},
  43(7):2193--2205, 2021.

\bibitem{li2018sh}
Fengqiang Li, Florian Willomitzer, Prasanna Rangarajan, Mohit Gupta, Andreas
  Velten, and Oliver Cossairt.
\newblock Sh-tof: Micro resolution time-of-flight imaging with superheterodyne
  interferometry.
\newblock In {\em 2018 IEEE International Conference on Computational
  Photography (ICCP)}, pages 1--10. IEEE, 2018.

\bibitem{Wyant15OSA}
James~C Wyant.
\newblock Interferometric optical metrology.
\newblock {\em OSA Century of Optics}, 2015.

\bibitem{BornWolf}
Max Born and Emil Wolf.
\newblock {Principles of optics: electromagnetic theory of propagation,
  interference and diffraction of light. }, Elsevier, 2013., p. 528.

\bibitem{Li:19}
Fengqiang Li, Florian Willomitzer, Prasanna Rangarajan, and Oliver Cossairt.
\newblock Mega-pixel time-of-flight imager with ghz modulation frequencies.
\newblock In {\em Imaging and Applied Optics 2019 (COSI, IS, MATH, pcAOP)},
  page CTh2A.2. Optical Society of America, 2019.

\bibitem{Wu20}
Yicheng Wu, Fengqiang Li, Florian Willomitzer, Ashok Veeraraghavan, and Oliver
  Cossairt.
\newblock Wished: Wavefront imaging sensor with high resolution and depth
  ranging.
\newblock In {\em 2020 IEEE International Conference on Computational
  Photography (ICCP)}. IEEE, 2020.

\bibitem{WilloCOSI19}
Florian Willomitzer, Fengqiang Li, Muralidhar~Madabhushi Balaji, Prasanna
  Rangarajan, and Oliver Cossairt.
\newblock High resolution non-line-of-sight imaging with superheterodyne remote
  digital holography.
\newblock In {\em Computational Optical Sensing and Imaging}, pages CM2A--2.
  Optical Society of America, 2019.

\bibitem{Willo19Arx}
Florian Willomitzer, Prasanna~V Rangarajan, Fengqiang Li, Muralidhar~M Balaji,
  Marc~P Christensen, and Oliver Cossairt.
\newblock Synthetic wavelength holography: An extension of gabor's holographic
  principle to imaging with scattered wavefronts.
\newblock {\em arXiv preprint arXiv:1912.11438}, 2019.

\bibitem{Willo21Nat}
Florian Willomitzer, Prasanna~V Rangarajan, Fengqiang Li, Muralidhar~M Balaji,
  Marc~P Christensen, and Oliver Cossairt.
\newblock Fast non-line-of-sight imaging with high-resolution and wide field of
  view using synthetic wavelength holography.
\newblock {\em Nature communications}, 12(1):1--11, 2021.

\bibitem{Cheng85}
Yeou-Yen Cheng and James~C Wyant.
\newblock Multiple-wavelength phase-shifting interferometry.
\newblock {\em Appl. Opt.}, 24(6):804--807, 1985.

\bibitem{polhemus73}
C~Polhemus.
\newblock Two-wavelength interferometry.
\newblock {\em Applied Optics}, 12(9):2071--2074, 1973.

\bibitem{tiziani1996dual}
HJ~Tiziani, A~Rothe, and N~Maier.
\newblock Dual-wavelength heterodyne differential interferometer for
  high-precision measurements of reflective aspherical surfaces and step
  heights.
\newblock {\em Applied optics}, 35(19):3525--3533, 1996.

\bibitem{PdG94}
Peter~J De~Groot.
\newblock Extending the unambiguous range of two-color interferometers.
\newblock {\em Applied optics}, 33(25):5948--5953, 1994.

\bibitem{falaggis2009multiwavelength}
Konstantinos Falaggis, David~P Towers, and Catherine~E Towers.
\newblock Multiwavelength interferometry: extended range metrology.
\newblock {\em Optics letters}, 34(7):950--952, 2009.

\bibitem{zhou2022review}
Haowen Zhou, Mallik~MR Hussain, and Partha~P Banerjee.
\newblock A review of the dual-wavelength technique for phase imaging and 3d
  topography.
\newblock {\em Light: Advanced Manufacturing}, 3(1):1--21, 2022.

\bibitem{PdG92}
Peter De~Groot and John McGarvey.
\newblock Chirped synthetic-wavelength interferometry.
\newblock {\em Optics letters}, 17(22):1626--1628, 1992.

\bibitem{GH04Enc}
G.~H{\"a}usler.
\newblock Speckle and coherence, Encyclopedia of Modern Optics, Elsevier,
  Academic Press., Oxford, pp 114-123, 2004.

\bibitem{GoodmanSpeck}
JW. Goodman.
\newblock {Speckle phenomena in optics: theory and applications }, Roberts and
  Company Publishers; 2007.

\bibitem{Willo21NatSupp}
Florian Willomitzer, Prasanna~V Rangarajan, Fengqiang Li, Muralidhar~M Balaji,
  Marc~P Christensen, and Oliver Cossairt.
\newblock \textit{Supplementary Material to:} fast non-line-of-sight imaging
  with high-resolution and wide field of view using synthetic wavelength
  holography.
\newblock {\em Nature communications}, 12(1):1--11, 2021.

\bibitem{HechtOptics}
Eugene Hecht.
\newblock {\em Optics}.
\newblock Pearson Education India, 2012.

\bibitem{huntley1993temporal}
Jonathan~M Huntley and Henrik Saldner.
\newblock Temporal phase-unwrapping algorithm for automated interferogram
  analysis.
\newblock {\em Applied Optics}, 32(17):3047--3052, 1993.

\bibitem{WilloDiss_19}
Florian Willomitzer.
\newblock {Single-Shot 3D Sensing Close to Physical Limits and Information
  Limits}, Dissertation, Springer Theses (2019).

\bibitem{wang2019one}
Kaiqiang Wang, Ying Li, Qian Kemao, Jianglei Di, and Jianlin Zhao.
\newblock One-step robust deep learning phase unwrapping.
\newblock {\em Optics express}, 27(10):15100--15115, 2019.

\bibitem{yin2019temporal}
Wei Yin, Qian Chen, Shijie Feng, Tianyang Tao, Lei Huang, Maciej Trusiak, Anand
  Asundi, and Chao Zuo.
\newblock Temporal phase unwrapping using deep learning.
\newblock {\em Scientific reports}, 9(1):1--12, 2019.

\bibitem{Dandliker:88}
R.~D\"{a}ndliker, R.~Thalmann, and D.~Prongu\'{e}.
\newblock Two-wavelength laser interferometry using superheterodyne detection.
\newblock {\em Opt. Lett.}, 13(5):339--341, May 1988.

\bibitem{Fercher:85}
A.~F. Fercher, H.~Z. Hu, and U.~Vry.
\newblock Rough surface interferometry with a two-wavelength heterodyne speckle
  interferometer.
\newblock {\em Appl. Opt.}, 24(14):2181--2188, Jul 1985.

\bibitem{Vry:86}
U.~Vry and A.~F. Fercher.
\newblock Higher-order statistical properties of speckle fields and their
  application to rough-surface interferometry.
\newblock {\em J. Opt. Soc. Am. A}, 3(7):988--1000, Jul 1986.

\bibitem{kotwal2022swept}
Alankar Kotwal, Anat Levin, and Ioannis Gkioulekas.
\newblock Swept-angle synthetic wavelength interferometry.
\newblock {\em arXiv preprint arXiv:2205.10655}, 2022.

\bibitem{kreis1997methods}
Thomas~M Kreis, Mike Adams, and Werner~PO J{\"u}ptner.
\newblock Methods of digital holography: a comparison.
\newblock In {\em Optical Inspection and Micromeasurements II}, volume 3098,
  pages 224--233. International Society for Optics and Photonics, 1997.

\bibitem{kim2010principles}
Myung~K Kim.
\newblock Principles and techniques of digital holographic microscopy.
\newblock {\em SPIE reviews}, 1(1):018005, 2010.

\bibitem{mann2008quantitative}
Christopher~J Mann, Philip~R Bingham, Vincent~C Paquit, and Kenneth~W Tobin.
\newblock Quantitative phase imaging by three-wavelength digital holography.
\newblock {\em Optics express}, 16(13):9753--9764, 2008.

\bibitem{fratz2021digital}
Markus Fratz, Tobias Seyler, Alexander Bertz, and Daniel Carl.
\newblock Digital holography in production: an overview.
\newblock {\em Light: Advanced Manufacturing}, 2(3):283--295, 2021.

\bibitem{fu2009dual}
Yu~Fu, Giancarlo Pedrini, Bryan~M Hennelly, Roger~M Groves, and Wolfgang Osten.
\newblock Dual-wavelength image-plane digital holography for dynamic
  measurement.
\newblock {\em Optics and Lasers in Engineering}, 47(5):552--557, 2009.

\bibitem{yamagiwa2018multicascade}
Masatomo Yamagiwa, Takeo Minamikawa, Cl{\'e}ment Trovato, Takayuki Ogawa, Dahi
  Ghareab~Abdelsalam Ibrahim, Yusuke Kawahito, Ryo Oe, Kyuki Shibuya, Takahiko
  Mizuno, Emmanuel Abraham, et~al.
\newblock Multicascade-linked synthetic wavelength digital holography using an
  optical-comb-referenced frequency synthesizer.
\newblock {\em Optics Express}, 26(20):26292--26306, 2018.

\bibitem{hase2021multicascade}
Eiji Hase, Yu~Tokizane, Masatomo Yamagiwa, Takeo Minamikawa, Hirotsugu
  Yamamoto, Isao Morohashi, and Takeshi Yasui.
\newblock Multicascade-linked synthetic-wavelength digital holography using a
  line-by-line spectral-shaped optical frequency comb.
\newblock {\em Optics Express}, 29(10):15772--15785, 2021.

\bibitem{javidi2005three}
Bahram Javidi, Pietro Ferraro, Seung-Hyun Hong, Sergio De~Nicola, Andrea
  Finizio, Domenico Alfieri, and Giovanni Pierattini.
\newblock Three-dimensional image fusion by use of multiwavelength digital
  holography.
\newblock {\em Optics letters}, 30(2):144--146, 2005.

\bibitem{FREUND199049}
Isaac Freund.
\newblock Looking through walls and around corners.
\newblock {\em Physica A: Statistical Mechanics and its Applications},
  168(1):49 -- 65, 1990.

\bibitem{WANG769}
L.~WANG, P.~P. HO, C.~LIU, G.~ZHANG, and R.~R. ALFANO.
\newblock Ballistic 2-d imaging through scattering walls using an ultrafast
  optical kerr gate.
\newblock {\em Science}, 253(5021):769--771, 1991.

\bibitem{2020NatRP...2..141Y}
Seokchan {Yoon}, Moonseok {Kim}, Mooseok {Jang}, Youngwoon {Choi}, Wonjun
  {Choi}, Sungsam {Kang}, and Wonshik {Choi}.
\newblock {Deep optical imaging within complex scattering media}.
\newblock {\em Nature Reviews Physics}, 2(3):141--158, February 2020.

\bibitem{Ntz_2010}
Vasilis Ntziachristos.
\newblock {Going deeper than microscopy: the optical imaging frontier in
  biology}.
\newblock {\em Nature Methods}, 7, 2010.

\bibitem{Dunsby_03}
C.~Dunsby and P.M.W. French.
\newblock {Techniques for depth-resolved imaging through turbid media including
  coherence-gated imaging}.
\newblock {\em J. Phys. D: Appl. Phys}, 36, 2003.

\bibitem{Hoshi_16}
Yoko~Hoshi M.D. and Yukio Yamada.
\newblock {Overview of diffuse optical tomography and its clinical
  applications}.
\newblock {\em Journal of Biomedical Optics}, 21(9):1 -- 11, 2016.

\bibitem{Yoo:s}
K.~M. Yoo and R.~R. Alfano.
\newblock Time-resolved coherent and incoherent components of forward light
  scattering in random media.
\newblock {\em Opt. Lett.}, 15(6):320--322, Mar.

\bibitem{Singh_14}
Alok~Kumar Singh, Dinesh~N. Naik, Giancarlo Pedrini, Mitsuo Takeda, and
  Wolfgang Osten.
\newblock Looking through a diffuser and around an opaque surface: A
  holographic approach.
\newblock {\em Opt. Express}, 22(7):7694--7701, Apr 2014.

\bibitem{Bertolotti_12}
Jacopo Bertolotti, Elbert Putten, Christian Blum, Ad~Lagendijk, Willem Vos, and
  Allard Mosk.
\newblock Non-invasive imaging through opaque scattering layers.
\newblock {\em Nature}, 491:232--4, 11 2012.

\bibitem{Mosk_12}
Allard Mosk, Ad~Lagendijk, Geoffroy Lerosey, and Mathias Fink.
\newblock Controlling waves in space and time for imaging and focusing in
  complex media.
\newblock {\em Nature Photonics}, 6:283--292, 05 2012.

\bibitem{Yaqoob_08}
Zahid Yaqoob, Demetri Psaltis, Michael Feld, and Changhuei Yang.
\newblock Optical phase conjugation for turbidity suppression in biological
  samples.
\newblock {\em Nature photonics}, 2:110--115, 02 2008.

\bibitem{vellekoop_10}
{Ivo Micha} Vellekoop, Aart Lagendijk, and Allard Mosk.
\newblock Exploiting disorder for perfect focusing.
\newblock {\em Nature photonics}, 4:320--322, 2010.
\newblock doi:10.1038/nphoton.2010.3.

\bibitem{Xu_11}
Xiao Xu, Honglin Liu, and Lihong Wang.
\newblock Time-reversed ultrasonically encoded optical focusing into scattering
  media.
\newblock {\em Nature photonics}, 5:154, 03 2011.

\bibitem{Doktofsky_20}
Daniel Doktofsky, Moriya Rosenfeld, and Ori Katz.
\newblock Acousto optic imaging beyond the acoustic diffraction limit using
  speckle decorrelation.
\newblock {\em Communications Physics}, 3:5, 12 2020.

\bibitem{Fac_NatPhys_19}
Daniele Faccio, Andreas Velten, and Gordon Wetzstein.
\newblock Non-line-of-sight imaging.
\newblock {\em Nature Reviews Physics}, 2:318--327, 2020.

\bibitem{Batarseh18}
M.~Batarseh, S.~Sukov, Z.~Shen, H~Gemar, R.~Rezvani, and A.~Dogariu.
\newblock Passive sensing around the corner using spatial coherence.
\newblock {\em Nature Communications 9, 3629 (2018)}.

\bibitem{OtooleNat_18}
Matthew O'Toole, David Lindell, and Gordon Wetzstein.
\newblock {Confocal non-line-of-sight imaging based on the light-cone
  transform}, Nature (2018).

\bibitem{MetzlerKeyhole}
C.~A. {Metzler}, D.~B. {Lindell}, and G.~{Wetzstein}.
\newblock Keyhole imaging:non-line-of-sight imaging and tracking of moving
  objects along a single optical path.
\newblock {\em IEEE Transactions on Computational Imaging}, 7:1--12, 2021.

\bibitem{LindellWave}
David~B. Lindell, Gordon Wetzstein, and Matthew O'Toole.
\newblock Wave-based non-line-of-sight imaging using fast f-k migration.
\newblock {\em ACM Trans. Graph. (SIGGRAPH)}, 38(4):116, 2019.

\bibitem{heide2014diffuse}
Felix Heide, Lei Xiao, Wolfgang Heidrich, and Matthias~B Hullin.
\newblock Diffuse mirrors: 3d reconstruction from diffuse indirect illumination
  using inexpensive time-of-flight sensors.
\newblock In {\em Proceedings of the IEEE Conference on Computer Vision and
  Pattern Recognition}, pages 3222--3229, 2014.

\bibitem{LiuNat_19}
Xiaochun Liu, Ibon Guillen, Marco~La Manna, Ji~Hyun Nam, Syed~Azer Reza,
  Toan~Huu Le, Adrian Jarabo, Diego Gutierrez, and Andreas Velten.
\newblock {Non-line-of-sight imaging using phasor-field virtual wave optics},
  Nature (2019).

\bibitem{Velten12}
Andreas Velten, Thomas Willwacher, Otkrist Gupta, Ashok Veeraraghavan, Moungi
  Bawendi, and Ramesh Raskar.
\newblock {Recovering three-dimensional shape around a corner using ultrafast
  time-of-flight imaging}, Nature Communications (2012).

\bibitem{Faccio:19}
Daniele Faccio.
\newblock Non-line-of-sight imaging.
\newblock {\em Opt. Photon. News}, 30(1):36--43, Jan 2019.

\bibitem{katz14}
Ori Katz, Pierre Heidmann, Mathias Fink, and Sylvain Gigan.
\newblock Non-invasive single-shot imaging through scattering layers and around
  corners via speckle correlations.
\newblock {\em Nature photonics}, 8(10):784, 2014.

\bibitem{katz2012looking}
Ori Katz, Eran Small, and Yaron Silberberg.
\newblock Looking around corners and through thin turbid layers in real time
  with scattered incoherent light.
\newblock {\em Nature Photonics}, 6:549{\textendash}553, 2012.

\bibitem{lindell20}
David~B Lindell and Gordon Wetzstein.
\newblock Three-dimensional imaging through scattering media based on confocal
  diffuse tomography.
\newblock {\em Nature Communications}, 11(1):1--8, 2020.

\bibitem{gariepy_16}
Genevieve Gariepy, Francesco Tonolini, Robert Henderson, Jonathan Leach, and
  Daniele Faccio.
\newblock Detection and tracking of moving objects hidden from view.
\newblock {\em Nature photonics}, 10, 2016.

\bibitem{SAVA2019527}
Paul Sava and Erik Asphaug.
\newblock Seismology on small planetary bodies by orbital laser doppler
  vibrometry.
\newblock {\em Advances in Space Research}, 64(2):527 -- 544, 2019.

\bibitem{Beer}
Beer.
\newblock Bestimmung der absorption des rothen lichts in farbigen
  fluessigkeiten.
\newblock {\em Annalen der Physik}, 162(5):78--88, 1852.

\bibitem{Freund_88}
Isaac Freund, M.~Rosenbluh, and Shechao Feng.
\newblock Memory effects in propagation of optical waves through disordered
  media.
\newblock {\em Physical review letters}, 61:2328--2331, 12 1988.

\bibitem{nam2021low}
Ji~Hyun Nam, Eric Brandt, Sebastian Bauer, Xiaochun Liu, Marco Renna, Alberto
  Tosi, Eftychios Sifakis, and Andreas Velten.
\newblock Low-latency time-of-flight non-line-of-sight imaging at 5 frames per
  second.
\newblock {\em Nature communications}, 12(1):1--10, 2021.

\bibitem{Edrei:16}
Eitan Edrei and Giuliano Scarcelli.
\newblock Optical imaging through dynamic turbid media using the fourier-domain
  shower-curtain effect.
\newblock {\em Optica}, 3(1):71--74, Jan 2016.

\bibitem{PrasannaSPIE}
Prasanna Rangarajan, Florian Willomitzer, Oliver Cossairt, and Marc~P.
  Christensen.
\newblock {Spatially resolved indirect imaging of objects beyond the line of
  sight}.
\newblock In Jean~J. Dolne, Mark~F. Spencer, and Markus~E. Testorf, editors,
  {\em Unconventional and Indirect Imaging, Image Reconstruction, and Wavefront
  Sensing 2019}, volume 11135, pages 124 -- 131. International Society for
  Optics and Photonics, SPIE, 2019.

\bibitem{Aparna_Corr}
Aparna Viswanath, Prasanna Rangarajan, Duncan MacFarlane, and Marc~P
  Christensen.
\newblock Indirect imaging using correlography.
\newblock In {\em Imaging and Applied Optics 2018 (3D, AO, AIO, COSI, DH, IS,
  LACSEA, LS\&C, MATH, pcAOP)}, page CM2E.3. Optical Society of America, 2018.

\bibitem{Balaji:17}
Muralidhar~Madabhushi Balaji, Prasanna Rangarajan, Duncan MacFarlane, Andreas
  Corliano, and Marc~P. Christensen.
\newblock Single-shot holography using scattering surfaces.
\newblock In {\em Imaging and Applied Optics 2017 (3D, AIO, COSI, IS, MATH,
  pcAOP)}, page CTu2B.1. Optical Society of America, 2017.

\bibitem{metzler2020deep}
Christopher~A Metzler, Felix Heide, Prasana Rangarajan, Muralidhar~Madabhushi
  Balaji, Aparna Viswanath, Ashok Veeraraghavan, and Richard~G Baraniuk.
\newblock Deep-inverse correlography: towards real-time high-resolution
  non-line-of-sight imaging.
\newblock {\em Optica}, 7(1):63--71, 2020.

\bibitem{reza2019phasor}
Syed~Azer Reza, Marco La~Manna, Sebastian Bauer, and Andreas Velten.
\newblock Phasor field waves: A huygens-like light transport model for
  non-line-of-sight imaging applications.
\newblock {\em Optics express}, 27(20):29380--29400, 2019.

\bibitem{gupta2015phasor}
Mohit Gupta, Shree~K Nayar, Matthias~B Hullin, and Jaime Martin.
\newblock Phasor imaging: A generalization of correlation-based time-of-flight
  imaging.
\newblock {\em ACM Transactions on Graphics (ToG)}, 34(5):1--18, 2015.

\bibitem{Rayleigh}
Lord Rayleigh, Philos. Mag. 8, 403 (1879). Reprinted in his Scientific Papers
  (Cambridge U. Press, 1899), Vol. 1, pp. 432-435.

\bibitem{hausler2011limitations}
Gerd H{\"a}usler and Svenja Ettl.
\newblock Limitations of optical 3d sensors.
\newblock In {\em Optical Measurement of Surface Topography}, pages 23--48.
  Springer, 2011.

\bibitem{Dresel:92}
Thomas Dresel, Gerd H\"{a}usler, and Holger Venzke.
\newblock Three-dimensional sensing of rough surfaces by coherence radar.
\newblock {\em Appl. Opt.}, 31(7):919--925, Mar 1992.

\bibitem{hausler1994range}
G~H{\"a}usler.
\newblock Range sensing of the first, the second, and the third kind.
\newblock In {\em Proc. of EOS TOPICAL MEETING Optical Metrology and
  Nanotechnology}, pages 27--30, 1994.

\bibitem{haeusler1993coherence}
Gerd H{\"a}usler and Jochen Neumann.
\newblock Coherence radar: an accurate 3d sensor for rough surfaces.
\newblock In {\em Optics, Illumination, and Image Sensing for Machine Vision
  VII}, volume 1822, pages 200--205. SPIE, 1993.

\bibitem{ettl1998roughness}
Peter Ettl, Berthold~E Schmidt, M~Schenk, Ildiko Laszlo, and Gerd H{\"a}usler.
\newblock Roughness parameters and surface deformation measured by coherence
  radar.
\newblock In {\em International Conference on Applied Optical Metrology},
  volume 3407, pages 133--140. SPIE, 1998.

\end{thebibliography}

\end{document}